\documentclass{article}
\pdfoutput=1
\usepackage{jcappub}
\usepackage{amsmath}
\usepackage{amssymb}
\usepackage{hyperref}
\usepackage{multirow}
\usepackage{soul}
\usepackage{bm}
\usepackage{caption}
\usepackage{comment}

\definecolor{darkred}{RGB}{175,0,0}

\definecolor{darkteal}{RGB}{0,111,111}

\definecolor{lightblue}{RGB}{100,149,237}

\definecolor{darkmagenta}{rgb}{0.55, 0.0, 0.55}

\title{Testing the growth of cosmic structures during the Dark Ages}

\author[a,b,c,d]{Elena Vanetti,}
\emailAdd{elena.vanetti@iac.es}

\author[a,b,e]{Eleonora Vanzan,}
\emailAdd{evanzan@tauex.tau.ac.il}

\author[a,b,f]{Nicola Bellomo,}
\emailAdd{nicola.bellomo@unipd.it}

\author[a,b,f]{Alvise Raccanelli}
\emailAdd{alvise.raccanelli.1@unipd.it}

\affiliation[a]{Dipartimento di Fisica e Astronomia G. Galilei, Università degli Studi di Padova, via Marzolo 8, I-35131 Padova, Italy}
\affiliation[b]{INFN, Sezione di Padova, Via Marzolo 8, I-35131, Padova, Italy}
\affiliation[c]{Instituto de Astrofísica de Canarias (IAC), Calle Vía Láctea, S/N, E-38205, La Laguna, Tenerife, Spain}
\affiliation[d]{Departamento de Astrofísica, Universidad de La Laguna, Avenida Francisco Sánchez, E-38205, La Laguna, Tenerife, Spain}
\affiliation[e]{School of Physics and Astronomy, Tel-Aviv University, Chaim Levanon St 55, Tel-Aviv 69978, Israel}
\affiliation[f]{INAF - Osservatorio Astronomico di Padova, Vicolo dell'Osservatorio 5, I-35122 Padova, Italy}

\abstract{
Hydrogen 21-cm Line Intensity Mapping offers the unique opportunity to access the Dark Ages and trace the formation and evolution of the large scale structure of the Universe prior to star and galaxy formation.
In this work we investigate the potential of future Earth- and Moon-based 21-cm surveys to constrain the growth of structures during the currently unexplored redshift range~$30 < z < 200$.
On the one hand we show how foreground contamination could limit the capabilities of future instruments in achieving precision below the~$10\%$ level. 
On the other hand, observations from the far side of the Moon have the potential to reach percent or even sub-percent precision in terms of reconstructing the growth of cosmic structures, if foregrounds are robustly accounted for. 
Such exquisite precision will provide tight constraints on models that induce deviations from~$\Lambda$CDM, not only during the Dark Ages, but also during recombination or that manifest mostly in the low-redshift Universe, like Early Dark Energy and nDGP models, respectively. 
Thus, because of their insensitivity to non-linearities or astrophysical processes, line intensity mapping surveys will provide a formidable consistency check to potential claims of discoveries of new physics that affect the growth of structures.
}

\begin{document}

\maketitle

\section{Introduction}

Structure formation in the cosmological standard model, the so-called~$\Lambda$CDM+General Relativity (GR), has been well understood for almost four decades~\cite{peebles:lssformation, blumenthal:lssformation, davis:lssformation}.
Initial (linear) perturbations in the matter field, whose origin can be traced back to the inflationary epoch, start growing after horizon re-entry because of gravitational instability, until they become nonlinear and form virialized structures~\cite{bernardeau:lssreview, cooray:halomodelreview}.
In the big picture of this paradigm the only required ingredients are the presence of primordial perturbations, cold dark matter (CDM), i.e.,~a non-relativistic species that does not interact with baryons, and gravity.

Several Cosmic Microwave Background (CMB) experiments~\cite{aghanim:planckcosmoparameters, aiola:actcosmoparameters, pan:sptcosmoparameters} have returned a clear picture of the Universe at high redshift~$(z \gtrsim 10^3)$, around the epoch of recombination.
In particular, these experiments clarified the role of CDM in providing the potential wells into which, later in cosmic history, baryons fall.
Similarly, galaxy redshift surveys at low redshift~$(z\lesssim 2)$ confirmed that the final outcome of the growth of structure, i.e.,~a cosmic web made of galaxies, filaments and voids, is fully compatible with~$\Lambda$CDM+GR predictions and the initial conditions for baryons measured at recombination~\cite{hawkins:2dfsurvey, tegmark:sdsssurvey, ross:2dfsdsssurvey, daangela:2dfsdsssurvey, guzzo:2dfvimossurvey, blake:wigglezsurvey, blake:gamasurvey, blake:wigglezsurveyII, beutler:6dfsurvey, sanchez:sdssiiisurvey, chuang:sdssiiisurvey, feix:sdsssurvey, okumura:fmossurvey, achitouv:6dfsurvey, alam:sdssiiisurvey, zhao:sdssivsurvey, nadathur:bosssurvey, aubert:sdssivsurvey, Samushia:2012, Raccanelli:2013, Huterer:2015, DESI:2024hhd, DESI:2024mwx}.
However, despite this overwhelming theoretical and experimental success, the theoretical foundations and deep explanations of the physics behind $\Lambda$CDM+GR are still lacking.
Moreover, the presence of a few tensions in our current set of observations regarding the expansion history and growth of structure could hint at the presence of beyond-$\Lambda$CDM Physics~\cite{heymans:sigma8tension, abbott:sigma8tension, verde:H0review, DESI:2024mwx}.

There are theoretical models attempting to both explain the observed data and the accelerated expansion of the Universe, and to cure these tensions by extending~$\Lambda$CDM, invoking the presence of additional species or deviations from GR, either in the early or late Universe, see, e.g.,~\cite{divalentino:H0solutions, schonenberg:H0olympics, khalife:H0solution}.
However, without a complete observational picture from recombination to the present day, none of these proposed models is able to offer a satisfactory explanation.
A novel and promising experimental probe is Line Intensity Mapping (LIM) (see e.g.,~\cite{kovetz:limstatusreport} for a review, and references therein).
LIM measures the integrated line emission of atomic or molecular spectral lines originating from unresolved sources, and, depending on the chosen spectral line, allows to bridge the gap between low- and high-redshift cosmological observables.
Given the large number of LIM experiments that are taking and will take place in the near future, see, e.g.,~refs.~\cite{kovetz:astrocosmolim, silva:limlss}, it is timely to investigate LIM potential to test the growth of structures.

Although several authors have already explored the possibility of using LIM to test deviations from~$\Lambda$CDM at redshift~$z\lesssim 6~$\cite{Masui:2009cj, Bull:2014rha, Karagiannis:2022ylq, Scott:2022fev, Casas:2022vik, Castorina:2019zho}, the technique appears promising even for studying the reionization and cosmic dawn epochs~\cite{MoradinezhadDizgah:2023src}.
However, in this work, we focus on the Hydrogen 21-cm line, which provides a rather unique access to the so-called ``Dark Ages'', in the~$30 \lesssim z \lesssim 200$ redshift window~\cite{field:21cmline, zaldarriaga:21cmline}.
The study of the Dark Ages is motivated by the presence of three significant features which make them the perfect target for high-precision cosmological measurements~\cite{Furlanetto:2019jso}: observables in this epoch \textit{(i)} are well described by linear theory, \textit{(ii)} provide access to a number of modes significantly larger than CMB experiments~\cite{cole:darkagesmodes} since Hydrogen perturbations are not affected by Silk damping after photon-baryon decoupling, and
\textit{(iii)} suffer from minimal astrophysical uncertainties since stars have not formed yet.

In this work we showcase the outstanding potential of 21-cm LIM in constraining the growth of cosmic structure during the Dark Ages, thus in bridging the gap between recombination and the late Universe. 
Hereafter, the term ``growth of structure'' indicates the growth of matter density fluctuations in the linear regime, representing the first stage of cosmic web formation.
We focus on both model-independent and model-dependent deviations from~$\Lambda$CDM, where in the latter case we consider theoretical models of particular interest for early and late Universe phenomenology, such as Early Dark Energy~\cite{Poulin:2023lkg, Kamionkowski:2022pkx} and Dvali-Gabadadze-Porrati~\cite{Dvali:2000hr} models, respectively.
These models differ from those investigated in ref.~\cite{dekruijf:21cmconstraints}, where the deviations are induced by early Universe physics, at the level of the primordial curvature power spectrum.
By contrasting Earth- and Moon-based experiments, we highlight the advantages of the latter in providing tighter constraints on the cosmological parameters associated to deviations from~$\Lambda$CDM, with improvements ranging from a factor few to an order of magnitude.
Moreover, if foregrounds are robustly accounted for, lunar experiments are able to reach percent and sub-percent precision on most models considered in this work, i.e.,~a precision competitive with that reached by complementary current and future cosmological experiments.

This article is organized as follows. 
In section~\ref{sec:dark_ages} we review the theoretical basis of 21-cm LIM and useful statistical forecasting tools.
In section~\ref{sec:alternative_models} we describe alternative models that generate deviations from the~$\Lambda$CDM growth of structure, while in section~\ref{sec:constraints_growth} we present the constraints on such models for future experiments. 
Finally, we conclude in section~\ref{sec:conclusions}.
Appendix~\ref{app:Cls21} reviews the derivation of the angular power spectrum of 21-cm brightness temperature fluctuations, appendix~\ref{app:resolution_with_foregrounds} contains a discussion about how to account for the foreground wedge in harmonic space, while appendix~\ref{app:tables_of_constraints} reports constraints corresponding to the figures presented in section~\ref{sec:constraints_growth}.


\section{Observing the Dark Ages}
\label{sec:dark_ages}


\subsection{Hydrogen 21-cm line}
\label{subsec:21cm}

The 21-cm Hydrogen spectral line corresponds to the hyperfine splitting of the ground state of neutral hydrogen into a singlet and triplet state~\cite{Furlanetto2006REVIEW, Pritchard:2011xb,Pen:2008fw,Chang:2010jp}. 
The energy difference between the two states is~$E_{21} = 5.9 \times 10^{-6} \, \text{eV}$, thus the emitted photon has a wavelength of~$\lambda_{21} = 21\ \mathrm{cm}$ or, equivalently, a frequency of $\nu_{21}=1420\ \text{MHz}$.
The relative occupation number density of neutral hydrogen atoms in the triplet~($n_1$) and singlet~($n_0$) states is encoded into the spin temperature~$T_s$ as
\begin{equation}
    \frac{n_1}{n_0} = 3 e^{-T_*/T_s}, 
\end{equation}
where~$T_* = E_{21}/k_B = 68\ \mathrm{mK}$, and~$k_B$ is the Boltzmann constant. 
The spin temperature is determined by competing collisional and radiative processes.
Collisions with free electrons, protons and neutral hydrogen atoms drive the spin temperature towards the Hydrogen kinetic temperature~$T_k$, while absorption of and stimulated emission induced by CMB photons drive~$T_s$ towards the CMB temperature~$T_{\gamma}$.
Therefore, we can describe the spin temperature as a weighted average of kinetic and CMB temperatures as~\cite{field:21cmline} 
\begin{equation}
    T^{-1}_s = \frac{T^{-1}_{\gamma} + x_k T^{-1}_k}{1+x_k}, \qquad x_k = \frac{T_*}{T_\gamma} \frac{C_{10}}{A_{10}},
\label{eq:Ts2}
\end{equation}
where~$C_{10}$ is the total collisional de-excitation rate~\cite{Kuhlen:2005cm}, and~$A_{10}$ is the spontaneous emission coefficient of the 21-cm transition.
Since the spin temperature is not directly observable, we aim to measure the 21-cm brightness temperature of neutral hydrogen in contrast with a background radio source (the CMB), which reads as
\begin{equation}
    T_{21} = \frac{T_s-T_\gamma}{(1+z)} (1-e^{-\tau_{21}}) \simeq \frac{T_s-T_\gamma}{(1+z)} \tau_{21} ,
\label{eq:T21z}
\end{equation}
where the optical depth of the 21-cm line is given by
\begin{equation}
    \tau_{21} = \frac{3 x_\mathrm{HI} n_H \lambda^3_{21}}{32\pi} \frac{E_{21}}{k_B T_s} \frac{A_{10}}{H(z)+(1+z) \partial_\parallel v_\parallel} \ll 1 ,
    \label{eq:tau}
\end{equation}
and~$x_\mathrm{HI}$ is the neutral Hydrogen fraction, $n_H$ is the Hydrogen number density, $H(z)$ is the Hubble expansion rate, and~$\partial_\parallel v_{\parallel}$ is the comoving gradient of the gas velocity along the line of sight.
During the Dark Ages, $T_{21}$ departs from zero around redshift~$z\sim 200$, when baryons thermally decouple from CMB photons and start cooling adiabatically, lowering the spin temperature to~$T_s\simeq T_k<T_{\gamma}$. 
Therefore, the 21-cm signal is observed in absorption, i.e.,~$T_{21}<0$.

In this work we focus on the statistics of 21-cm brightness temperature fluctuations around the global time-dependent background value~$\bar{T}_{21}$, specifically on their two-point function in harmonic space.
First we divide the total observational frequency band~$\Delta \nu_\mathrm{tot}$ into $N_\nu$~frequency bins of width~$\Delta \nu$ centered at the observed frequency~$\nu_i=\nu_{21}/(1+z_i)$, where~$z_i$ is the photon emission redshift.
The angular power spectrum of the 21-cm brightness temperature fluctuations during the Dark Ages is well approximated by~\cite{Pillepich:2006fj, Lewis:2007kz, AliHaimoud2013}
\begin{equation}
    C^{ij}_\ell = 4 \pi \int d\log k\ \mathcal{P}_\zeta(k) \mathcal{T}^{21}_\ell (k,\nu_i) \mathcal{T}^{21}_\ell(k,\nu_j),
    \label{eq:Cls21}
\end{equation}
where~$(i,j)$ label two arbitrary frequency channels, $k$'s are Fourier modes, and~$\mathcal{P}_\zeta$ is the almost scale-invariant primordial curvature power spectrum.
The 21-cm harmonic transfer function reads as
\begin{equation}
    \mathcal{T}^{21}_{\ell}(k,\nu_i) = \int dr W_{\nu_i}(r) \left[\alpha(r) j_\ell(kr) - f(r) \bar{T}_{21}(r) \partial^2_{kr} j_\ell(kr)\right] \mathcal{T}_b(k,r),
    \label{eq:21transfer}
\end{equation}
where~$r$ is the comoving distance, $j_\ell$ and~$\partial^2_{kr} j_\ell$ are Bessel functions and their second derivatives, respectively, $\alpha(r)$ is a time-dependent fitting function,~$f(r)=d\log\delta_m / d\log a$ is the linear growth rate of matter density perturbations, $\mathcal{T}_b(k,r)$ is the baryon transfer function, and~$W_{\nu_i}$ is a Gaussian window function centered at~$r_i=r(z_i)$ with variance~$\sigma_i$ set by the frequency bin width as
\begin{equation}
    \sigma_i = \frac{c(1+z_i)^2}{\nu_{21} H(z_i)} \frac{\Delta \nu}{2},
\end{equation}
where~$c$ is the speed of light.
This choice of comoving distance bin width ensures that frequency cross-bin angular power spectra are negligible because the signal is dominated by purely local effects~\cite{Lewis:2007kz, Munoz2015}. 
The interested reader will find more details on the derivation of equation~\eqref{eq:Cls21} in appendix~\ref{app:Cls21}.


\subsection{Forecasting the sensitivity of 21-cm experiments}
\label{subsec:forecasting_sensitivity}

The Fisher matrix analysis represents a convenient tool to forecast the precision of a hypothetical future experiment because it returns the best achievable error, i.e.,~the Cram\'er-Rao bound~\cite{fisher:fishermatrix, bunn:fishermatrix, vogeley:fishermatrix, tegmark:fishermatrix}.
Under the assumption of Gaussian-distributed data, absence of coupling between multipoles, and absence of correlation between frequency bins, the exact Fisher matrix element~\cite{Bellomo2020} corresponding to the pair of cosmological parameters~$\{\theta_\alpha,\theta_\beta\}$ reduces to
\begin{equation}
    F_{\alpha\beta} = f_\mathrm{sky} \sum_{i=1}^{N_\nu} \sum_{\ell=2}^{\ell_{i,\mathrm{max}}} \sigma^{-2}_{C^{ii}_\ell} \frac{\partial C^{ii}_\ell}{\partial \theta_\alpha} \frac{\partial C^{ii}_\ell}{\partial \theta_\beta}, 
\label{eq:fisher_matrix}
\end{equation}
where the parameter~$f_\mathrm{sky}$ accounts for a partial coverage of the sky, the frequency-dependent multipole~$\ell_{i,\mathrm{max}}$ defines the maximum resolution of the experiment in a given frequency channel, and the angular power spectra covariance reads as
\begin{equation}
    \sigma^2_{C^{ii}_\ell} = \frac{2}{2\ell+1} \left( C^{ii}_\ell + N_{i,\ell} \right)^2,
\end{equation}
where~$N_{i,\ell}$ is the angular power spectrum of the noise in the~$i$-th frequency channel.
In the case of radio interferometers with uniformly distributed antennas, the noise angular power spectrum can be estimated as~\cite{zaldarriaga:21cmline} 
\begin{equation}
   N_{i,\ell} = \frac{(2\pi)^3}{t_\mathrm{obs} \Delta\nu} \left( \frac{T_{i,\mathrm{sys}}}{f_\mathrm{cover} \ell_{i,\mathrm{max}}} \right)^2 ,
\end{equation}
where~$t_{\text{obs}}$ is the total observation time, $f_\mathrm{cover}$ is the fraction of interferometer area covered with antennas, and~$T_{i,\mathrm{sys}}$ is the system temperature. 
The system temperature is dominated by galactic synchrotron radiation in every frequency channel of interest, thus we can approximate it as~\cite{Furlanetto2006REVIEW} 
\begin{equation}
    T_{i,\mathrm{sys}} \approx 180 \left( \frac{\nu_i}{180\ \mathrm{MHz}} \right)^{-2.6}\ \mathrm{K},
\end{equation}
where both the amplitude and the spectral index are compatible with recent EDGES measurements~\cite{mozden:edgesskytemperature}.\footnote{This sky temperature is representative of observations performed at night, in a patch of sky that did not include the galactic center.
Measurements of sky temperature of different patches or performed at different times return a larger value, see, e.g.,~refs.~\cite{mozden:edgesskytemperature, rogers:skytemperature}, therefore the observation time does not necessarily reflect the ``human'' time.}
The maximum observable multipole is
\begin{equation}
   \ell_{i,\mathrm{max}} = \frac{2\pi D_\mathrm{base}}{\lambda_i} ,
\label{eq:maximum_multipole_interferometer}
\end{equation}
where~$D_\mathrm{base}$ is the largest baseline of the interferometer, and $\lambda_i = \lambda_{21}(1+z_i)$ is the redshifted 21-cm wavelength.
Finally, in order to understand the advantages of the tomographic approach, it is convenient to define a cumulative Fisher matrix as
\begin{equation}
    F^{(N)}_{\alpha\beta} = f_\mathrm{sky} \sum_{i=1}^{N} \sum_{\ell=2}^{\ell_{i,\mathrm{max}}} \sigma^{-2}_{C^{ii}_\ell} \frac{\partial C^{ii}_\ell}{\partial \theta_\alpha} \frac{\partial C^{ii}_\ell}{\partial \theta_\beta}, 
\label{eq:cumulative_fisher_matrix}
\end{equation}
where we consider the contribution only of the first~$N<N_\nu$ frequency channels.
When~$N=N_\nu$ we recover the Fisher matrix element of equation~\eqref{eq:fisher_matrix}.

The cosmological parameters we consider in this work are divided into two broad classes, $\bm{\theta}_{\Lambda\mathrm{CDM}}$ and~$\bm{\theta}_\mathrm{growth}$, where the former contains the common core~$\Lambda$CDM parameters while the latter contains the case-dependent parameters responsible for a growth of structure that deviates from~$\Lambda$CDM, see section~\ref{sec:alternative_models}.
Explicitly, the~$\Lambda$CDM parameters considered in this work are
\begin{equation}
    \bm{\theta}_{\Lambda\mathrm{CDM}} = \left\lbrace 100\theta_s, \omega_\mathrm{cdm}, \omega_\mathrm{b}, \log\left( 10^{10}A_s \right), n_s \right\rbrace,
\end{equation}
where~$\theta_s$ is the angular size of the sound horizon at recombination, $\omega_\mathrm{cdm}$ and~$\omega_\mathrm{b}$ are the cold dark matter and baryon physical densities, respectively, $A_s$ and~$n_s$ are the amplitude and tilt of the almost scale-invariant primordial curvature perturbation power spectrum, respectively.
Ultimately, the errors on each cosmological parameter~$\theta_\alpha$ are provided by the diagonal elements of the inverse Fisher matrix defined in equation~\eqref{eq:fisher_matrix} as~$\sigma_{\theta_\alpha} = \sqrt{\left(F^{-1}\right)_{\alpha\alpha}}$.
Similarly, the error coming from the cumulative Fisher approach in equation~\eqref{eq:cumulative_fisher_matrix} is given by~$\sigma^{(N)}_{\theta_\alpha} = \sqrt{\left((F^{N})^{-1}\right)_{\alpha\alpha}}$.
Therefore it is convenient to introduce the cumulative error ratio
\begin{equation}
    \mathcal{C}_{\theta_\alpha} = \frac{\sigma^{(N)}_{\theta_\alpha}}{\sigma_{\theta_\alpha}},
\label{eq:cumulative_error_ratio}
\end{equation}
which quantifies how quickly we saturate the error obtained including all frequency bins, i.e., whether the additional frequency channels included in the analysis are noise- or signal-dominated.

Foregrounds represent a significant challenge for~21-cm cosmology, since they are~$10^5$ times larger than the signal we aim to constrain~\cite{liu:dataanalysis21cm}.
While synchrotron emission represents the main foreground, there are other (subdominant) foregrounds sourced for instance by free-free emission or galactic/extragalactic radio sources, that still dominate over the 21-cm signal.
While in theory these smooth foregrounds can be subtracted from sky maps~\cite{deoliveiracosta:21cmforegrounds, liu:21cmforegrounds, zheng:21cmforegrounds}, the presence of instrumental systematics considerably limits this option, see, e.g.,~refs.~\cite{opperman:foregroundremoval, switzer:foregroundremoval, wolleben:foregroundremoval}.
Moreover, the presence of the ionosphere and radio frequency interference severely limits the exploitation of low frequency channels for an Earth-based experiment~\cite{vedantham:groundobservationlimitationI, datta:groundobservationlimitations, vedantham:groundobservationlimitationsII, vedantham:groundobservationlimitationsIII}, thus the idea of designing a Moon-based observatory~\cite{lazio:lunarradioarray}.

In this work we exploit the fact that many of these foregrounds, including those generated by the Earth’s atmosphere, appear to be comparable with thermal noise only in the so-called ``foreground wedge''~\cite{datta:foregroundwedge, vedantham:foregroundwedge, morales:foregroundwedge, parsons:foregroundwedge, trott:foregroundwedge, hazelton:foregroundwedge, pober:foregroundwedge, liu:foregroundwedgeI, liu:foregroundwedgeII, Seo:foregroundwedge}.
The foreground wedge is a section of the~$(k_\perp,k_\parallel)$ 2D Fourier space defined in each frequency channel by
\begin{equation}
    k_\parallel \lesssim \theta_0 d_A(z_i) H(z_i) k_\perp,
\label{eq:foreground_wedge}
\end{equation}
where~$k_\parallel$ and~$k_\perp$ are the vectors parallel and across the line-of-sight, respectively, $\theta_0$ is the field-of-view angle, and~$d_A$ is the angular diameter distance.
Since the two-point function in harmonic space receives contributions from all~$k$ modes, we limit our analysis to those multipoles where contamination from foregrounds is kept at a reasonable minimum. 
Inspired by ref.~\cite{pober:foregroundwedgesize}, we consider both a ``conservative'' and ``optimistic'' scenario with respect to the perspective of foreground removal.
In the former we consider contamination from systematics to be significant across the entire sky, i.e.,~$\theta_{0}^\mathrm{cons} = \pi$, while in the latter we assume them to be restrained to the primary beam, i.e.,~$\theta_{0}^\mathrm{opt} = 0.2$~\cite{braun:skasensitivity, macario:skasensitivity}.
In terms of maximum multipoles, we have that the conservative and optimistic scenarios correspond to
\begin{equation}
    \ell^\mathrm{cons}_{i,\mathrm{for}} \lesssim 0.14 \frac{\nu_i}{\delta \nu}, \qquad \ell^\mathrm{opt}_{i,\mathrm{for}} \lesssim  2.24\frac{\nu_i}{\delta \nu},
\label{eq:maximum_foreground_multipole}
\end{equation}
respectively, where~$\delta \nu$ is the spectral resolution of the measurement.
We refer the reader to appendix~\ref{app:resolution_with_foregrounds} for the derivation of equation~\eqref{eq:maximum_foreground_multipole}.

At the practical level, we consider four different experimental setups, including one state-of-the-art Earth-based experiment and three futuristic Moon-based observatories~\cite{bernal:smbhseeds, short:dmdarkages}.
\begin{itemize}
    \item \textbf{aSKAO}, an advanced version of the Square Kilometer Array Observatory able to probe the end of the Dark Ages at redshift~$z \approx 30$. 
    This observatory has~$N_\nu = 6$ frequency channels~$\Delta\nu_\mathrm{tot}=[40.3,46.3]\ \mathrm{MHz}$ with a bandwidth of~$\Delta\nu=1\ \mathrm{MHz}$, a baseline of~$D_\mathrm{base}=100\ \mathrm{km}$, a cover fraction of~$f_\mathrm{cover}=0.2$, and it observes the entirety of the sky, i.e.,~$f_\mathrm{sky}=0.75$, for~$t_\mathrm{obs}=5\ \mathrm{yr}$.

    \item \textbf{LRA1}, a Lunar Radio Array on the far side of the Moon.
    It has the same bandwidth, fraction of observed sky, and observation time as aSKAO, but it has a baseline of~$D_\mathrm{base}=30\ \mathrm{km}$, a cover fraction of~$f_\mathrm{cover}=0.1$, and it observes the totality of the Dark Ages by having~$N_\nu = 40$ frequency channels in the~$\Delta\nu_\mathrm{tot}=[6.3,46.3]\ \mathrm{MHz}$ frequency band.

    \item \textbf{LRA2}, a second, less conservative experiment, having the same specifics of LRA1 but with a baseline of~$D_\mathrm{base}=100\ \mathrm{km}$ and a cover fraction of~$f_\mathrm{cover}=0.2$, as aSKAO.

    \item \textbf{LRA3}, a third Lunar Radio Array representing the ideal Moon-based interferometer. 
    We extend the baseline to~$D_\mathrm{base}=300\ \mathrm{km}$ and increase the cover fraction to~$f_\mathrm{cover}=0.75$.
\end{itemize}

The indicative resolution of the experiments is of order~$\ell_\mathrm{max}(z=30) \simeq \mathcal{O}(10^5)$ and~$\ell_\mathrm{max}(z=200) \simeq \mathcal{O}(10^4)$ for~$D_{\rm{base}}\sim 10^2\ \mathrm{km}$, decreasing by one order of magnitude for~$D_{\rm{base}}\sim 10\ \mathrm{km}$. 
Regarding the foreground wedge, we assume a spectral resolution of~$\delta\nu = 10\ \mathrm{kHz}$~\cite{liu:dataanalysis21cm, mcquinn:21cmforecast, bowman:21cmforecast}.
Therefore, since~$\ell_{i,\mathrm{for}}^{\mathrm{cons}} \simeq \mathcal{O}(10^2)$, we find that, in the conservative foreground removal scenario, excluding the foreground wedge from our analysis imposes a maximum multipole orders of magnitude smaller than that associated with the interferometer resolution. 
Instead, in the optimistic case, a smaller range of modes is lost to foreground removal since~$\ell_{i,\mathrm{for}}^{\mathrm{opt}}\simeq \mathcal{O}(10^4)$. This resolution limitation highlights the need for developing efficient foreground-removal techniques to exploit the full potential of 21-cm Line Intensity Mapping.


\section{Alternative models of structure growth}
\label{sec:alternative_models}

Since nonlinear effects and uncertainties connected to stellar evolution do not affect the 21-cm signal in the Dark Ages, we can use it as a ``clean'' probe of structure formation.
We aim to test the capabilities of both Earth- and Moon-based experiments in constraining deviations from pure~$\Lambda$CDM+GR, using both a model-dependent and -independent approach.
All models considered in this work reproduce the~$\Lambda$CDM background evolution history and have the same inflationary initial conditions.
Deviations arise only at the level of growth of perturbations, either because of a modification of General Relativity or due to the presence of additional species.
Some of the scenarios investigate deviations from~$\Lambda$CDM happening during the Dark Ages themselves; others introduce new effects mainly during recombination or the post-reionization era, while still leaving a clear imprint also on the 21-cm signal.
Remarkably, 21-cm experiments are sensitive enough to constrain also the latter classes of models.
In all cases presented in this section, we employ modified versions of the public Boltzmann solver code \texttt{CLASS}~\cite{Blas:2011rf} to compute the baryon transfer function, the growth rate, and other relevant quantities entering into the calculation of the angular power spectra defined in equation~\eqref{eq:Cls21}.


\subsection{Model-independent approach on reconstructing growth history}

Matter density perturbations~$\delta_m$ grow at a rate proportional to the scale factor~$a$ at all scales during the matter-dominated era.
Therefore, it is convenient to describe the growth of perturbation in terms of the linear growth rate function~$f = d\log\delta_m / d\log a$.
Following the approach of ref.~\cite{Amendola:2016saw}, we model the growth rate as a step function which takes values~$f_i=f(z_i)$ in each frequency channel.
Each~$f_i$ is treated as an independent parameter in the forecast, thus in this case~$\bm{\theta}_\mathrm{growth}=\{f_1, ..., f_{N_\nu} \}$, with fiducial values taken from the publicly available code~\texttt{CLASS}~\cite{Blas:2011rf}.
The fiducial values of the core~$\Lambda$CDM parameters are taken from the \textit{Planck} 2018 (TT,TE,EE + low-E) analysis and they read as~\cite{aghanim:planckcosmoparameters}
\begin{equation}
    \bm{\theta}_{\Lambda\mathrm{CDM}} = \{ 100\theta_s = 1.04109, \omega_\mathrm{cdm} = 0.1202, \omega_\mathrm{b} = 0.02236, \log(10^{10}A_s)= 3.045, n_s=0.9649 \}.
\label{eq:lcdm_fiducial_reconstruction}
\end{equation}


\subsection{Deviations arising during the Dark Ages}

Although many different General Relativity extensions have been explored during the past decades, see, e.g.,~ref.~\cite{ade:planckmodifiedgravity, Durrer:2008in} and refs. therein, here we implement an approach based on the~``$\mu-\eta$'' parametrization~\cite{amendola:mgparametrization, zhao:muetaparametrization}. 
In this framework, the growth of structure at the linear level is altered by introducing two free functions of time and scale,~$\mu(k,z)$ and~$\eta(k,z)$, that effectively change the value of Newton's constant~$G$ and the relation between the Bardeen potentials~$\Psi$ and~$\Phi$ as
\begin{equation}
    k^2\Psi(k,z) = -4\pi G a^2 \mu(k,z) \bar{\rho} \Delta, \qquad \Phi(k,z) = \eta(k,z) \Psi(k,z),
\label{eq:mu_eta_parametrization}
\end{equation}
where~$\bar{\rho}$ is the background energy density, and~$\Delta$ is the gauge-invariant energy density contrast.
In this work we consider
\begin{equation}
    \mu(k,z) = 1 + (\mu_0-1) \mathcal{S}(k,z), \qquad \eta(k,z) = 1 + (\eta_0-1) \mathcal{S}(k,z),
\end{equation}
where the~$\Lambda$CDM limit is recovered at all scales and redshifts when~$\mu_0=\eta_0=1$.
The ``switch'' function~$\mathcal{S}$ is chosen in such a way that before and after the Dark Ages, and at small scales, we recover the~$\Lambda$CDM limit even when~$\mu_0,\eta_0 \neq 1$.
Inspired by refs.~\cite{zhao:muetaparametrization, alonso:screening, spuriomancini:screening, bosi:gwxlss}, we define
\begin{equation}
    \mathcal{S}(k,z) = \frac{1}{4} \left[ 1 - \tanh \left( \frac{z-z_{\rm on}}{\Delta z_{\rm on}} \right) \right] \left[ 1 - \tanh \left( \frac{z_{\rm off}-z}{\Delta z_{\rm off}} \right) \right] e^{-\frac{1}{2} \left( k/k_* \right)^2},
\end{equation}
where~$z_\mathrm{on},z_\mathrm{off}$ are two transition redshifts at which deviations are switched on and off, $\Delta z_\mathrm{on}, \Delta z_\mathrm{off}$ describe how fast the transition between the two regimes happens, and $k_*$ is a reference scale above which deviations are gradually switched off.
This scale dependence is inspired by, but not necessarily connected to, screening scenarios in Modified Gravity models~\cite{bellini:eftofde, khoury:chameleonI, khoury:chameleonII, hinterbichler:symmetron, brax:dilaton, babichev:kmouflage, vainshtein:vainshtein}.
When~$z_\mathrm{off}\lesssim z \lesssim z_\mathrm{on}$ and~$k \lesssim k_*$ we have~$\mathcal{S}(k,z)\simeq 1$, thus~$\mu \simeq \mu_0$ and~$\eta \simeq \eta_0$, while in the opposite regimes~$\mathcal{S}(k,z) \simeq 0$ and we recover the General Relativity limit.

We set~$z_\mathrm{on}=500, z_\mathrm{off}=20$ and~$\Delta z_\mathrm{on}=50, \Delta z_\mathrm{off}=4$  as reference values for the transition redshifts and widths, respectively.
Regarding the reference scale, we choose~$k_*=10^{-2}, 10^{-1}, 1\ \mathrm{Mpc}^{-1}$, as well as a scale-independent case where~$k_* \rightarrow \infty$.
The fiducial values for the core cosmological parameters are those of equation~\eqref{eq:lcdm_fiducial_reconstruction}, while the fiducial values for the growth parameters are~$\bm{\theta}_\mathrm{growth} = \{ \mu_0=1, \eta_0=1 \}$.


\subsection{Late universe phenomenology}
\label{subsec:nDGP_theory}
In this section we showcase the potential of 21-cm observations to constrain models that induce the bulk of their deviations from~$\Lambda$CDM after the Dark Ages ended.
One example of this class of theories is the Dvali-Gabadadze-Porrati (DGP) model~\cite{Dvali:2000hr}, where deviations from General Relativity are a consequence of the propagation of gravity across an extra fifth dimension.
The weakness of gravity is due to its propagation in a 5D Minkowski space, while our Universe remains embedded in 4D brane, i.e.,~the known 4D gravitational interaction is recovered only at short distances on the 4D brane.
This theory has two branches of homogeneous and isotropic solutions, denoted as ``normal'' and ``self-accelerating''. 
Here we focus on the normal branch of the theory (nDGP), since the self-accelerating branch is plagued by ghost instabilities~\cite{Gregory:2007xy, koyama:mudgpmodel}.
This class of models is a 1-parameter extension of~$\Lambda$CDM, where the extra parameter is the crossover scale~$r_c$ between the~5D and~4D phenomenology of the theory.

In this work we consider the nDGP model proposed in ref.~\cite{Schmidt:2009sv}, where the~$\Lambda$CDM expansion history is left unchanged but the growth of cosmic structure is altered.
Growth of perturbations in nDGP can be conveniently remapped into the $\mu-\eta$ parametrization defined in equation~\eqref{eq:mu_eta_parametrization} as~\cite{lue:mudgpmodel, koyama:mudgpmodel}
\begin{equation}
    \mu(z) = 1 + \frac{1}{3\beta(z)}, \qquad \eta(z) = \frac{1-\frac{1}{3\beta(z)}}{1+\frac{1}{3\beta(z)}}, \qquad \beta(z) = 1 + \frac{H}{H_0\sqrt{\Omega_\mathrm{rc}}}\left(1+\frac{\dot{H}}{3H^2}\right),
\end{equation}
where~$H_0$ is the Hubble expansion rate today, $\dot{H}$ is the Hubble expansion rate derivative with respect to cosmic time, and the parameter~$\Omega^{-1}_\mathrm{rc} = 4 r_c^2 H_0^2$ quantifies the size of the crossover scale compared to that of the Hubble horizon today.
When the crossover scale is much larger than the cosmological horizon, i.e.,~when~$r_c H(z) \gg 1$, we have~$\beta \to \infty$, thus~$\mu,\eta\to 1$ and deviations from General Relativity are relegated to scales that cannot be probed by cosmological experiments.
Even though during the Dark Ages nDGP models are responsible for a percent-level difference in the growth factor with respect to~$\Lambda$CDM, these small deviations can still be captured by future 21-cm observations.
Also in this case we consider as fiducial values for the cosmological parameters those appearing in equation~\eqref{eq:lcdm_fiducial_reconstruction}, while we take~$\bm{\theta}_\mathrm{growth} = \{ \Omega_\mathrm{rc}=0.2 \}$ as fiducial value for the extra nDGP parameter, motivated by current upper bounds~\cite{Piga:2022mge}.


\subsection{Deviations arising before the Dark Ages}
\label{subsec:ede_theory}

Finally, we also consider Early Dark Energy (EDE) models, which change the growth of structure before the beginning of the Dark Ages, leaving, however, a potentially detectable imprint in the 21-cm signal.
EDE has been largely studied as a possible solution to the Hubble tension~\cite{Karwal:2016vyq, Poulin:2018cxd, Lin:2019qug, Smith:2019ihp, sobotka:2024tat}, since it induces a decrease in the size of the comoving sound horizon around recombination by increasing the Hubble expansion rate.
Its phenomenology can be explained by the presence of an additional fluid with relative abundance of order~$10\%$ around matter-radiation equality, and equation of state such that~$w_\mathrm{EDE}(z\gtrsim 3300) \simeq -1$ and~$w_\mathrm{EDE}(z\lesssim 3300) \gtrsim 1/3$. 
In general, EDE models require a dark matter abundance larger than the~$\Lambda$CDM one to avoid the decay of gravitational potentials around the time of recombination, the so-called ``early ISW'' effect, thus the growth of perturbations is naturally enhanced in this class of models~\cite{Poulin:2023lkg}.

A typical realization of EDE is achieved via an ultra-light self-interacting scalar field frozen in place by Hubble friction before recombination.
Well-known candidates that display this phenomenology are axion-like particles~\cite{Marsh:2011gr, Hlozek:2014lca, Marsh:2015xka} with potential 
\begin{equation}
    V(\phi_a) = m^2_a f^2_a \left( 1-\cos\frac{\phi_a}{f_a} \right)^n,
\end{equation}
where the scalar field~$\phi_a$ has mass~$m_a \sim 10^{-27}\ \mathrm{eV}$, $f_a$ is a characteristic energy scale, and we consider~$n=3$ to avoid additional fine tuning of the potential~\cite{Hill:2020osr}.
This choice of the exponent also agrees with the most likely value found by Cosmic Microwave Background, Baryon Acoustic Oscillations and Supernovae data joint-analysis~\cite{Poulin:2023lkg}.
Instead of directly working with parameters appearing in the Lagrangian, it is convenient to use the alternative set of parameters~$\{f_{\rm EDE}, \log_{10} z_c, \theta_\mathrm{ini}\}$, where~$f_{\rm EDE}$ is the ratio between EDE density and the critical one evaluated at critical redshift~$z_c$, at which EDE becomes dynamical, and~$\theta_\mathrm{ini} = \phi_{a,\mathrm{ini}}/f_a$ is the normalized scalar field initial value.
We use~\texttt{CLASS\_EDE}~\cite{Hill:2020osr}, a publicly available modified version of~\texttt{CLASS}, to estimate EDE effects on matter power spectra.
In this case, fiducial values for the cosmological and growth parameters are~$\bm{\theta}_{\Lambda\mathrm{CDM}} = \{ 100\theta_s = 1.04152, \omega_\mathrm{cdm} = 0.1306, \omega_\mathrm{b} = 0.02253, \log(10^{10}A_s) = 3.098, n_s = 0.9889 \}$ and~$\bm{\theta}_\mathrm{growth} = \{ f_{\rm EDE} = 0.122, \log_{10} z_c = 3.562, \theta_\mathrm{ini} = 2.83 \}$, respectively~\cite{Smith:2019ihp}.


\section{Constraints on the growth of structure} 
\label{sec:constraints_growth}

\begin{figure}[ht]
    \centerline{
    \includegraphics[width=\columnwidth]{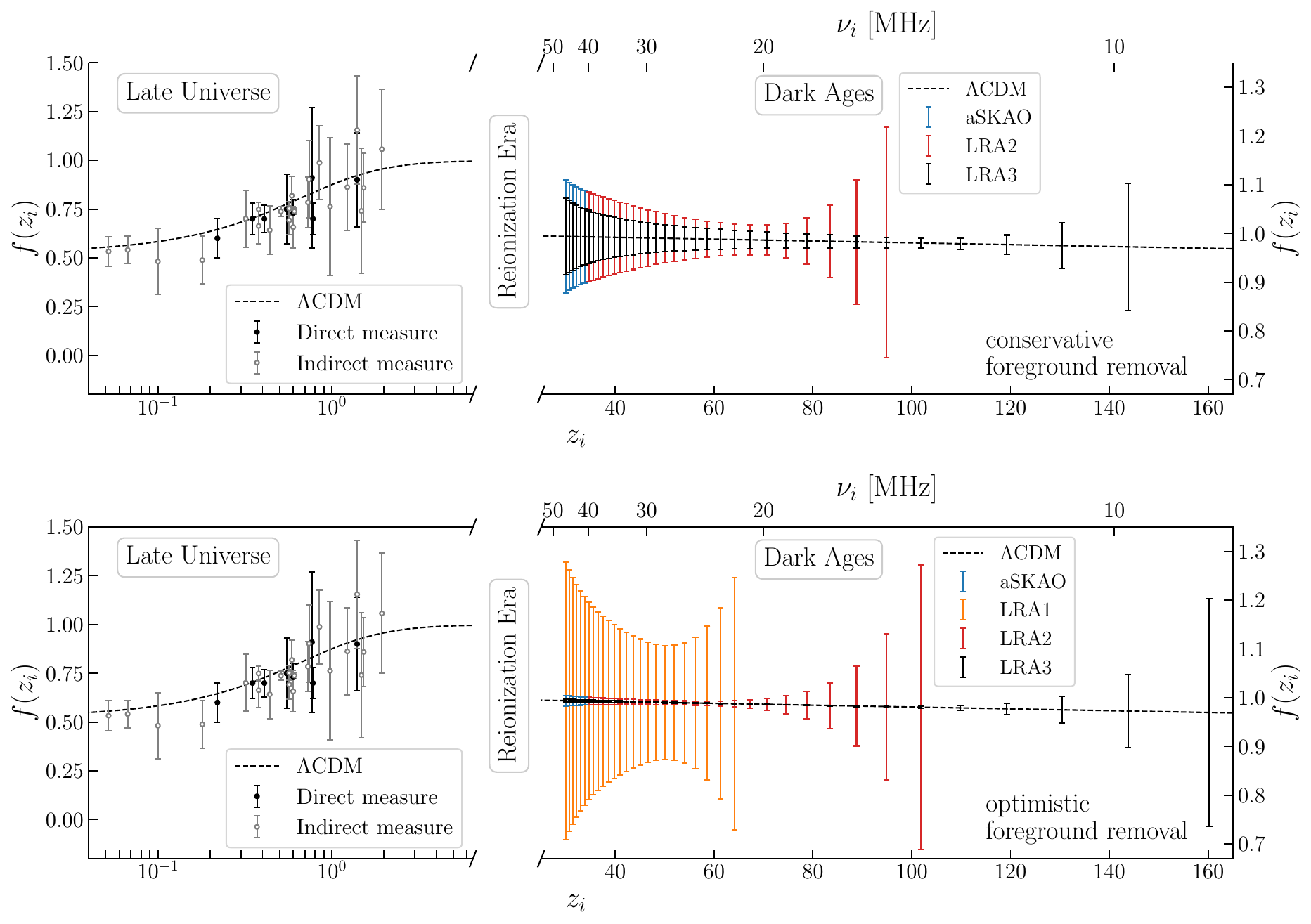}}
    \caption{Direct (\textit{black}) and indirect (\textit{gray}) measures of the growth rate in the late Universe~\cite{hawkins:2dfsurvey, tegmark:sdsssurvey, ross:2dfsdsssurvey, daangela:2dfsdsssurvey, guzzo:2dfvimossurvey, blake:wigglezsurvey, blake:gamasurvey, blake:wigglezsurveyII, beutler:6dfsurvey, Samushia:2012, sanchez:sdssiiisurvey, chuang:sdssiiisurvey, feix:sdsssurvey, okumura:fmossurvey, achitouv:6dfsurvey, alam:sdssiiisurvey, zhao:sdssivsurvey, nadathur:bosssurvey, aubert:sdssivsurvey}, $\Lambda$CDM theoretical values \textit{(black dashed line)}, and predicted sensitivity for individual frequency channels of aSKAO (\textit{blue}), LRA1 (\textit{orange}) LRA2 (\textit{red}) and LRA3 (\textit{black}) with conservative (\textit{top panel}) and optimistic (\textit{lower panel}) foreground removal.
    In the case of Dark Ages experiments we report constraints only for those frequency channels where a precision of order~$30\%$ is reached.
    For the sake of clarity, note that the left and right y-axis scales do not match.}
\label{fig:fz_reconstruction}
\end{figure}
Galaxy redshift surveys provide a large number of independent measurements of the growth history at redshift~$z \lesssim 2$ by targeting different kinds of tracers (LRG, ELG, Ly$\alpha$, voids, etc.)~\cite{hawkins:2dfsurvey, tegmark:sdsssurvey, ross:2dfsdsssurvey, daangela:2dfsdsssurvey, guzzo:2dfvimossurvey, blake:wigglezsurvey, blake:gamasurvey, blake:wigglezsurveyII, beutler:6dfsurvey, Samushia:2012, sanchez:sdssiiisurvey, chuang:sdssiiisurvey, feix:sdsssurvey, okumura:fmossurvey, achitouv:6dfsurvey, alam:sdssiiisurvey, zhao:sdssivsurvey, nadathur:bosssurvey, aubert:sdssivsurvey}.
These growth rate measurements are either ``direct'' or ``indirect'', where the former means that individual~$f(z_i)$s are estimated from data, while in the latter case the measured quantity is~$f(z)\sigma_8(z)$, where~$\sigma_8(z)=D(z)\sigma_8$, $D(z)$ is the linear growth factor normalized to unity at redshift~$z=0$, and~$\sigma_8$ is the linear matter field fluctuation RMS today, see, e.g.,~ref.~\cite{Huterer:2015, avila:growthrateoverview} for an overview.
Similarly, measuring the 21-cm line at different frequencies allows to investigate structure formation over a wide range of redshifts and reconstruct the neutral hydrogen distribution via a tomographic analysis.

We show in figure~\ref{fig:fz_reconstruction} the present and expected (at the moment of this work) future landscape of~$f(z)$ measurements.
Late Universe indirect measurements are converted into growth rate estimates by imposing a Planck prior on~$\sigma_8$~\cite{aghanim:planckcosmoparameters}.
The high-redshift late Universe ($z \gtrsim 2$) is still an uncharted epoch, however future experiments such as~SPHEREx\cite{dore:spherexwhitepaperI, dore:spherexwhitepaperII, dore:spherexwhitepaperIII}, Roman~\cite{Spergel2015:Roman}, and the proposed MegaMapper~\cite{Schlegel:2019eqc, Schlegel:2022vrv} and SIRMOS~\cite{SIRMOS}, will be able to explore even that redshift range.
We were unable to find constraints on the growth rate during the reionization era, between redshift~$6.5 \lesssim z \lesssim 30$, however we note that the Hydrogen 21-cm line signal is still present in that epoch, even though in that case it is also affected by star and galaxy formation processes~\cite{Furlanetto2006REVIEW, Pritchard:2011xb}.
Finally, we present the forecast on the growth rate for our four experimental setups.
We observe that, even in the conservative foreground removal scenario, lunar experiment configurations with a baseline of order~$D_{\rm{base}}\sim10^2$, achieve a precision below the~$10\%$ level over a remarkably broad redshift range, namely~$33<z_i\lesssim 85$ and~$30<z_i\lesssim 130$ for LRA2 and LRA3, respectively. Outside these ranges, errors rapidly increase because synchrotron radiation noise dominates over the clustering signal at low frequencies, thus we do not display those results. As a consequence, in the optimistic foreground removal scenario, the aforementioned ranges extend only slightly to~$30 < z_i \lesssim 90$ for LRA2 and~$30 < z_i \lesssim 140$ for LRA3. However, in this case, sub-percent precision is achievable for~$30 < z_i \lesssim 65$ (LRA2) and~$30< z_i \lesssim 110$ (LRA3).
This sensitivity allows us to test even models that affect the growth of structure mostly outside of the Dark Ages.
At last, the smaller baseline and cover fraction of LRA1 limit its constraining capability to~$10-30\%$ in the redshift range~$30<z\lesssim65$ in the optimistic case, and to remain above the $45\%$ level in the conservative one. 
Overall, we note that the maximum sensitivity is typically reached around frequencies probing~$z_i \approx 50-60$, where the amplitude of the angular power spectrum of 21-cm fluctuations reaches its maximum.
\begin{figure}[ht] 
    \centerline{
    \includegraphics[width=\columnwidth]{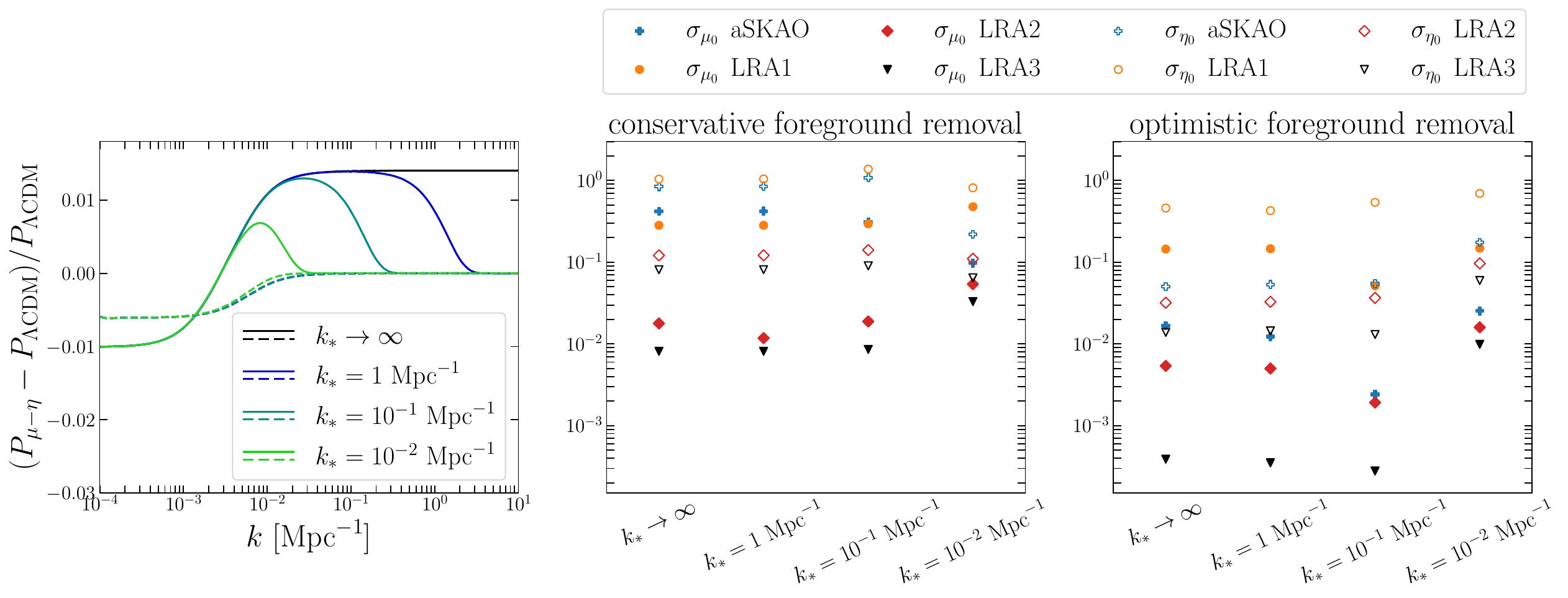}}
    \caption{\textit{Left panel:} $\mu-\eta$ matter power spectrum relative deviation with respect to~$\Lambda$CDM+GR prediction at redshift~$z=30$.
    Different scale-dependent deviations are induced by individual~$0.5\%$ changes from unity in the~$\mu_0$ (\textit{solid lines}) or~$\eta_0$ (\textit{dashed lines}) parameters, while keeping the other ones fixed at their~$\Lambda$CDM+GR values. 
    \textit{Center and right panel:} marginalized errors on~$\mu_0$ (\textit{filled markers}) and~$\eta_0$ (\textit{empty markers}) for aSKAO (\textit{blue}), LRA1 (\textit{orange}), LRA2 (\textit{red}), and LRA3 (\textit{black}) with conservative and optimistic foreground removal. The fiducial values of the parameters used for the Fisher forecast are~$\{\mu_0=1, \eta_0=1\}$;~$\Lambda$CDM parameters as listed in equation~\eqref{eq:lcdm_fiducial_reconstruction}.}
\label{fig:mPk_errors_mueta}
\end{figure}

In terms of model-dependent deviations from~$\Lambda$CDM during the Dark Ages, we show in the left panel of figure~\ref{fig:mPk_errors_mueta} the effects of changing the evolution of the Bardeen potentials on the matter power spectrum and, in the center and right panels, the constraints we find on the~$\mu_0,\eta_0$ parameters.
The left panel reports the relative difference between the matter power spectrum when varying~$\mu_0$ or~$\eta_0$ by~$0.5\%$ from unity for different choices of~$k_*$.
While~$\mu$ alters in opposite fashions the growth of structure at scales above and below the cosmological horizon~$k_\mathrm{hor} \simeq aH \approx 10^{-3}\ \mathrm{Mpc}^{-1}$, $\eta$ induces appreciable changes only on super-horizon scales and on a restricted range of the largest sub-horizon scales, thus we expect constraints on this parameter to be less tight.
The chosen values of~$k_*$ reflect those scales our experiments will be mostly sensitive to.
Already at this point, we note that~$k_*$ choices in the~$\eta$ function have little to no impact in altering the matter power spectrum at redshift~$z=30$.

We observe in the center and right panels of figure~\ref{fig:mPk_errors_mueta} the benefits of a tomographic approach, especially with conservative foreground removal. In this case, Moon-based experiments yield constraints that are typically an order of magnitude tighter than those coming from the ground, when baselines are comparable. 
When comparing the optimistic and conservative foreground removal scenarios for a fixed instrumental configuration, in most instances of interest, errors increase by at least one order of magnitude for aSKAO but only by a factor of~$3$ or less for LRA configurations, highlighting how tomography makes constraints less sensitive to the amount of modes lost to the foreground wedge. 
Moreover, while constraints on~$\mu_0$ depend on the choice of~$k_*$, constraints on~$\eta_0$ do not, because it only affects the growth of the largest sub-horizon modes.
Finally, when the growth of perturbations differs only on large scales, as in the~$k_*=10^{-2}\ \mathrm{Mpc}^{-1}$ case, constraints become less tight because of the~$\mu_0-\eta_0$ degeneracy due to the fact that they induce a similar scale-dependent suppression/enhancement on the matter power spectrum.
We report in table~\ref{tab:mu0eta0_constraints} of appendix~\ref{app:tables_of_constraints} the marginalized errors displayed in the center and right panels of figure~\ref{fig:mPk_errors_mueta}.
Additionally, we explicitly check that marginalized errors do not change by more than a factor of a few when varying the transition speeds~$\Delta z_\mathrm{on}, \Delta z_\mathrm{off}$.

\begin{figure}[ht]
    \centerline{
    \includegraphics[width=0.87\columnwidth]{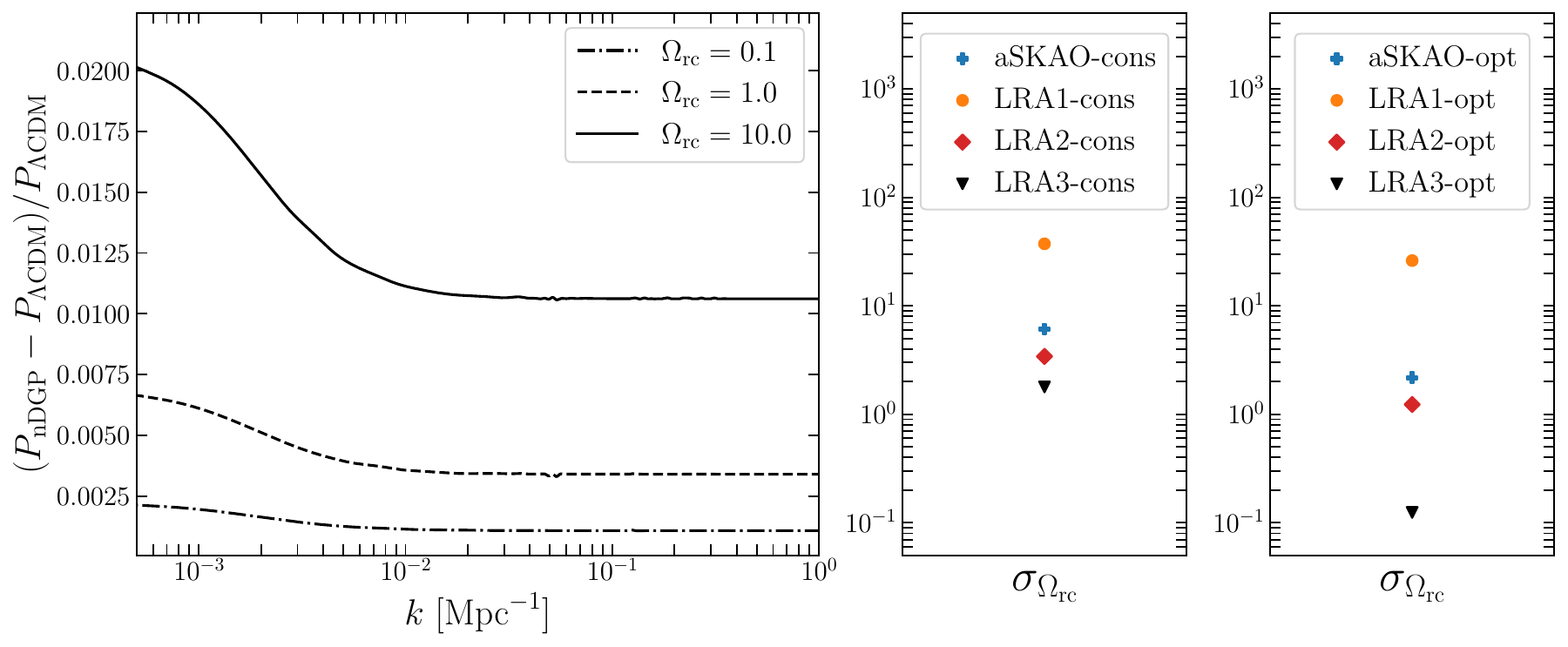}}
    \caption{\textit{Left panel:} nDGP matter power spectrum relative deviation with respect to~$\Lambda$CDM prediction at redshift~$z=30$. 
    \textit{Center and right panel:} marginalized errors on~$\Omega_\mathrm{rc}$ for aSKAO (\textit{blue}), LRA1 (\textit{orange}), LRA2 (\textit{red}) and LRA3 (\textit{black}) with conservative and optimistic foreground removal. Fiducial values of the parameters used for the Fisher  forecast:~$\{\Omega_{\rm{rc}}=0.2\}$;~$\Lambda$CDM parameters as listed in equation~\ref{eq:lcdm_fiducial_reconstruction}.}
\label{fig:mPk_errors_nDGP}
\end{figure}

Thanks to their remarkable sensitivity, 21-cm Moon-based experiments also have the potential to provide a complementary test of theories that significantly differ from~$\Lambda$CDM only at late times.
The nDGP model induces deviations in the matter power spectrum at all scales, as can be seen in the left panel of figure~\ref{fig:mPk_errors_nDGP}.
The scale-dependence is exclusively due to the fact that sub- and super-horizon modes grow at a different pace when~$\{\mu,\eta\} \neq 1$, as we already illustrated in the previous scenario. 
We observe in the center and right panels of figure~\ref{fig:mPk_errors_nDGP} that most experimental configurations examined in this work have no constraining power over ~$\Omega_\mathrm{rc}$, even when considering optimistic foreground removal.  Only an experiment with instrumental specifics comparable to those of LRA3 is able to constrain~$\Omega_\mathrm{rc}$ down to the~$60\%$ precision level in the optimistic case and therefore provide competitive and complementary constraints on this class of theories with respect to existing~\cite{Raccanelli:2013, Piga:2022mge} and future~\cite{bosi:gwxlss, frusciante:nDGPeuclidforecast, bose:nDGPeuclidforecast} surveys.
Also in this case we report in table~\ref{tab:nDGP_constraints} of appendix~\ref{app:tables_of_constraints} the values of marginalized errors displayed in the center and right panels of figure~\ref{fig:mPk_errors_nDGP}. Moreover, we find that marginalized relative errors are largely unaffected by one order of magnitude changes in the fiducial value of $\Omega_{\rm{rc}}$.

\begin{figure}[ht]
    \centerline{
    \includegraphics[width=\columnwidth]{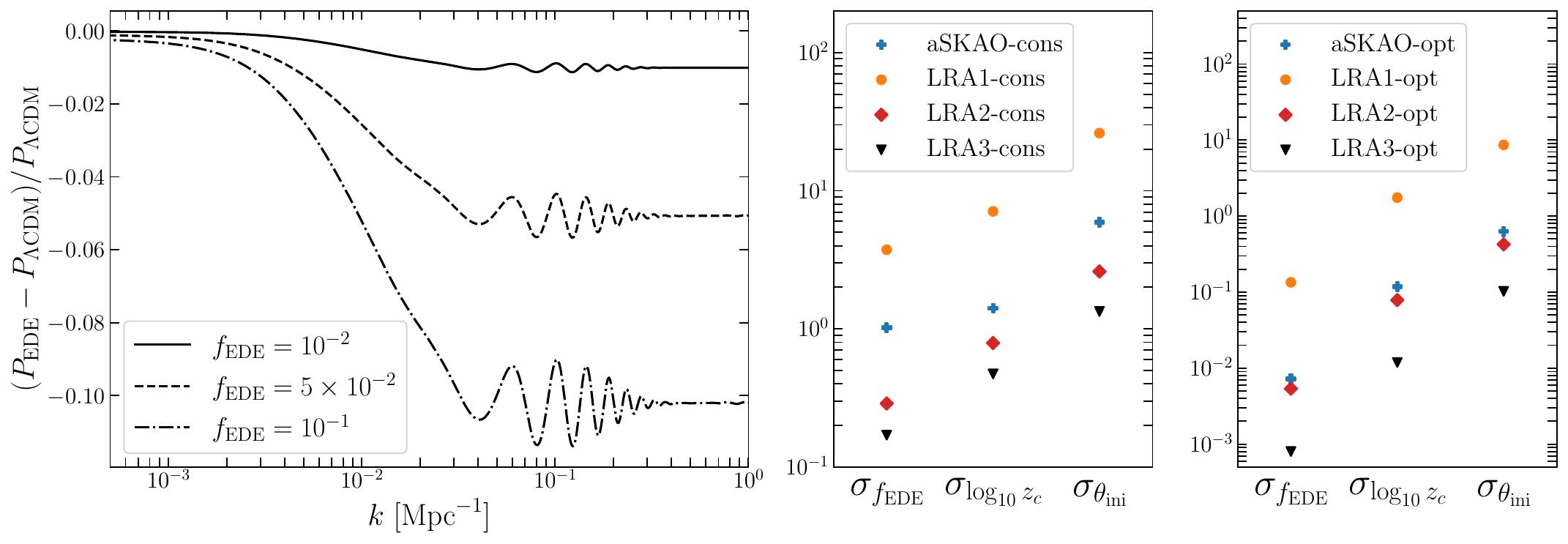}}
    \caption{\textit{Left panel:} EDE matter power spectrum relative deviation with respect to~$\Lambda$CDM prediction at redshift~$z=30$ for different EDE abundance. 
    Cosmological parameter for~$\Lambda$CDM are given in equation~\eqref{eq:lcdm_fiducial_reconstruction}.
    EDE fixed parameters are~$\{\log_{10} z_c = 3.562, \theta_\mathrm{ini} = 2.83 \}$.
    \textit{Center and right panel:} marginalized errors on EDE model parameters for aSKAO (\textit{blue}), LRA1 (\textit{orange}), LRA2 (\textit{red}) and LRA3 (\textit{black}) with conservative and optimistic foreground removal. Fiducial values of the parameters used for the Fisher forecast:~$\{ f_{\rm EDE} = 0.122, \log_{10} z_c = 3.562, \theta_\mathrm{ini} = 2.83 \}$;~$\Lambda$CDM parameters as listed in section~\ref{subsec:ede_theory}.}
\label{fig:mPk_errors_EDE}
\end{figure}

Following a similar logic, we turn our attention to the Early Dark Energy (EDE) scenario, where deviations from~$\Lambda$CDM take place considerably before the onset of the Dark Ages, around the time of matter-radiation equality.
Because of the decay of gravitational potentials due to EDE, the growth of matter perturbations is suppressed when compared to~$\Lambda$CDM.
Thus, we expect such a model to leave an imprint that is potentially detectable even at lower redshift despite the EDE fluid itself decays away faster than radiation, as we show in the left panel of figure~\ref{fig:mPk_errors_EDE}. 
In the conservative scenario, foreground removal highly limits the constraining capability on EDE parameters for all instrumental setups, since the majority of small scales where the growth is damped by EDE are removed with the wedge. 
In the optimistic scenario, aSKAO and LRA2 constraints on EDE parameters are compatible with existing constraints found by complementary datasets (\textit{Planck} TT,TE,EE + lensing, BAO, Pantheon SNIa, SHOES), see, e.g.,~refs.~\cite{Smith:2019ihp, Poulin:2023lkg, McDonough:2023qcu}. 
Finally, we find that the LRA3 configuration can improve constraints on~$f_{\rm{EDE}}$ and~$\log_{10} z_c$ by at least an order of magnitude, and enhance the precision on~$\theta_{\rm{ini}}$ by a factor of a few. 
In particular, given that solving the Hubble tension requires an abundance of EDE of order~$f_\mathrm{EDE}\simeq 10^{-1}$, all examined configurations except LRA1 can detect the presence of EDE-induced effects (for such abundances) at more than~$5\sigma$ level in the optimistic case, potentially providing a definitive answer regarding the viability of this candidate as a solution to the Hubble tension.
As before, we report in table~\ref{tab:EDE_constraints} of appendix~\ref{app:tables_of_constraints} the values of marginalized errors displayed in the center and right panels of figure~\ref{fig:mPk_errors_nDGP}.

Finally, we analyze the gain in constraining power given by a tomographic approach.
In particular, in figures~\ref{fig:cumulative errors_cons} and~\ref{fig:cumulative errors_opt} of appendix~\ref{app:tables_of_constraints} we compare the marginalized error obtained by using only the first~$N$ frequency channels (and so redshift bins) with that obtained using all of them.
Despite being severely limited by foregrounds, it appears that Earth-based experiment would still benefit from having a handful more frequency channels, even though this kind of consideration has to be validated by more accurate predictions of the effects of systematics.
Conversely, we note that in LRA experiments, most of the information is contained in the first twenty channels, with marginal gains obtained by pushing tomography to even lower frequencies (higher redshifts), where thermal noise completely dominates the anisotropic clustering signal.

Constraints reported in this section are marginalized over the set of core cosmological parameters.
We find that in the vast majority of cases marginalization does not affect the magnitude of New Physics parameters by more than a factor of a few, preserving the order of magnitude of the errors.
Also, by the time these experiments start observations, the set of core cosmological parameters will already be measured with a very high degree of precision and accuracy, effectively mimicking the effect of marginalization.


\section{Conclusions}
\label{sec:conclusions}

Detecting anisotropies in the 21-cm brightness temperature of neutral Hydrogen will allow to reconstruct not only its spatial distribution, but also to trace structure formation across a currently unexplored redshift range. 
In this work we investigated the potential of future Earth- and Moon-based 21-cm Line Intensity Mapping (LIM) surveys to constrain the growth of cosmic structures, both in a model -dependent and -independent fashion.
In particular, we focused on the Dark Ages because of their complementarity with existing cosmological observations and their insensitivity to uncertainties related to the development of non-linear structures and astrophysical processes. We produce forecasts for both Earth-based surveys, which are able to reach only the end of the Dark Ages at $z\sim30$, and for a futuristic Moon-based instrument that will observe the entirety of the Dark Ages and avoid foregrounds generated by the Earth's atmosphere.
There is not yet a consensus on the efficiency of foreground removal in the frequency range of interest; therefore, we present results for both a pessimistic scenario, employing a conservative foreground removal strategy, and an optimistic one that showcases the full potentialities of 21-cm LIM.

Our results show that Moon-based 21-cm LIM surveys with a baseline of order~$\sim 10^2$ km reach at minimum a ten percent precision level on the growth rate measurements over an extended redshift range during the Dark Ages.
The sensitivity of future 21-cm LIM surveys will enable high-precision constraints on theories that predict deviations from~$\Lambda$CDM at different moments of cosmic history, during and even outside the Dark Ages themselves. 
In particular, we focus on the nDGP model, which introduces deviations from General Relativity mostly at low redshifts (even though its percent-level difference in the growth factor is still observable during the Dark Ages) in order to explain the accelerated expansion, and on the EDE model, which alleviates the Hubble tension by introducing a new component behaving like Dark Energy that becomes relevant before recombination.
Even though most of the constraining power resides at lower redshifts, accessible also to Earth-based surveys, Moon-based experiments still provide tighter constraints, with precision improvements ranging from a factor of a few up to an order of magnitude. Lunar configurations, in fact, can take advantage of their higher redshift reach to mitigate the effect of the loss of modes due to foreground removal, and to break degeneracies between cosmological parameters.
We showed that, under optimistic foreground removal, lunar experiments will provide constraints competitive with future galaxy surveys and gravitational waves datasets for nDGP models, and will either confirm or disprove at more than~$5\sigma$ level the viability of EDE as a solution to the Hubble tension.
The strength of these experiments does not reside only in their constraining power on cosmological parameters, but also in their complementarity with current and future observations since they will test an epoch of the Universe currently unexplored.

We explore how LIM measurements during the Dark Ages can help bridge the gap in understanding the growth of cosmic structure between the epoch of recombination and the late Universe.
Future galaxy and LIM surveys have the potential to cover not only the redshift range corresponding to the high-redshift Universe, but also that corresponding to the epoch of reionization and cosmic dawn, providing unique insights on cosmology.
We envision that cross-correlating signals coming from different epochs will provide very stringent tests on multiple beyond-$\Lambda$CDM models, and eventually it will either confirm or rule out several new physics candidates that are currently adopted to solve cosmological tensions.
In conclusion, despite current technological limitations, future 21-cm LIM surveys will allow cosmology to enter a new era of precision measurements.


\acknowledgments
The authors would like to thank Andrea Begnoni, Jessie de Kruijf, Sarah Libanore, and Gabriele Perna for useful discussions.
Elena Vanetti acknowledges partial financial support by ASI Grant No. 2016-24-H.0. The project that gave rise to these results received the support of a fellowship from the a ``Caixa” Foundation (ID B006373). The fellowship code is LCF/BQ/DI24/12070001.
Nicola Bellomo is supported by PRD/ARPE 2022 ``Cosmology with Gravitational waves and Large Scale Structure - CosmoGraLSS''. 
AR acknowledges funding from the Italian Ministry of University and Research (MUR) through the ``Dipartimenti di eccellenza'' project ``Science of the Universe''.


\appendix
\section{The angular power spectrum of 21-cm brightness temperature fluctuations}
\label{app:Cls21}

The purpose of this appendix is to keep track of some of the steps taken to derive equation~\eqref{eq:Cls21}.
As recently pointed out in ref.~\cite{venumadhav:21cmlinefinitewidth}, the finite width of the 21-cm line requires the definition of the spin temperature to be changed to
\begin{equation}
    T_s^{-1} = \frac{x_\mathrm{CMB} T^{-1}_\gamma + x_k T^{-1}_k}{x_\mathrm{CMB}+x_k},
\end{equation}
where~$x_\mathrm{CMB} = \tau_{21}^{-1} \left(1 - e^{-\tau_{21}} \right)$.
However, during the Dark Ages we are consistently in the optically-thin regime, since between redshift~$30$ and~$200$ we have~$\tau_{21} \in [0.02,0.05]$, approximately.
Thus, the correction to the 21-cm brightness temperature induced by the finite width of the line is estimated to be of order~$\mathcal{O}(\tau_{21}/2)$, i.e.,~$1-2.5\%$~\cite{venumadhav:21cmlinefinitewidth}.
Although precision measurements of the 21-cm temperature will require an equally careful estimation of the theoretical signal, we choose for the moment to neglect these percent effect because we are not in possession of an equally accurate knowledge of the expected level of noise.

Fluctuations in the 21-cm brightness temperature are commonly approximated as (see, e.g., appendix B of ref.~\cite{Pillepich:2006fj})
\begin{equation}
   \delta T_{21} \simeq f_H \delta_H +f_{T_k} \delta_{T_k} + \bar{T}_{21} \delta_v,
\label{eq:T21pert}
\end{equation}
where~$\delta_H$ and~$\delta_{T_k}$ are the neutral hydrogen number density and kinetic temperature, respectively, the velocity perturbation~$\delta_v$ reads as
\begin{equation}
    \delta_v = - \frac{1+z}{H(z)} \partial_\parallel v_\parallel,
\end{equation}
and~$f_H,f_{T_k}$ are time-dependent functions given, for instance, in appendix~B of ref.~\cite{Pillepich:2006fj}.
We neglect fluctuations in the CMB temperature and ionization fraction in our derivation, since they have been shown to be subdominant in ref.~\cite{Lewis:2007kz}.

In this work we follow the approach developed in ref.~\cite{AliHaimoud2013} to connect baryon temperature and velocity fluctuations to the baryon overdensity perturbation.
Regarding the former, it was noted that under the approximation that baryons track dark matter evolution, thus~$\delta_b \propto a$, it is possible to recast the temperature perturbation as~$\delta_{T_k}(t) = C(t) \delta_b(t)$.
The functional form of~$C(t)$ is derived by solving the perturbed first law of thermodynamics
\begin{equation}
    \dot{\delta}_{T_k} - \frac{2}{3} \dot{\delta}_b + \Gamma_{\rm{C}} \frac{T_\gamma}{T_k} \delta_{T_k} = 0,
\end{equation}
with initial conditions~$\delta_{T_k,\mathrm{ini}} = 0$, where~$\Gamma_{\rm{C}}$ is the Compton scattering rate, and dots indicate derivatives with respect to cosmic time. 
Moreover, on top of this analytical argument, we also numerically check with~\texttt{CLASS} that such function~$C$ has a negligible scale dependence in the range of scales of interest for this work.
On the other hand, the velocity perturbation can be recast in terms of the baryon overdensity fluctuation using the continuity equation~$\dot{\delta}_b + a^{-1}\theta_b=0$, where~$\theta_b$ is the baryon velocity divergence, thus~$\theta_b=-aHf_b \delta_b$, where~$f_b$ is the growth rate of baryon density fluctuations, defined as~$f_b=d\log\delta_b/d\log a$. 
For the sake of facilitating a comparison with traditional large-scale structure growth rate measurements, we assume~$f_b\simeq f$.
In reality, especially at the beginning of the Dark Ages, $f$ \textit{underestimates} the real value of~$f_b$, since baryons have not already caught up with dark matter, especially at small scales; thus, with our forecast we obtain \textit{more conservative}, i.e., larger, errors.
On the other hand, we verified that towards the end of the Dark Ages, where most of the constraining power of our experiments reside (see, e.g., figures~\ref{fig:cumulative errors_cons} and~\ref{fig:cumulative errors_opt}), constraints on~$f$ are fundamentally unaffected by this assumption.
Finally, since~$\delta_H \sim \delta_b$, we arrive at the following expression in Fourier space
\begin{equation}
    \delta T_{21}(\textbf{k},r(z)) = \left[ \alpha(z) + \mu^2 f(z) \bar{T}_{21}(z) \right] \delta_b(\textbf{k},z), 
    \label{eq:dT21k}
\end{equation}
where~$\alpha = f_H + C f_{T_k}$, and~$\mu$ is the cosine of the angle between the Fourier mode~$\textbf{k}$ and the line-of-sight~$\mathbf{r}$. 
The spherical harmonic coefficients, as per common practice, are obtained as
\begin{equation}
    a_{\ell m}^i = \int d\Omega_{\hat{\mathbf{r}}} Y^*_{\ell m}(\hat{\mathbf{r}}) \times \int dr W_{\nu_i}(r) \int\frac{d^3k}{(2\pi)^3} e^{i\mathbf{k}\cdot\mathbf{r}} \delta T_{21}(\textbf{k},r).
\end{equation}
Equation~\eqref{eq:Cls21} is then easily obtained by taking the expectation value of pairs of spherical harmonic coefficients, where derivatives of Bessel functions appear by recasting~$\mu^2$ in terms of derivatives of~$e^{i\mathbf{k}\cdot\mathbf{r}}$. 

The authors of ref.~\cite{Lewis:2007kz} correctly pointed out that the RHS of equation~\eqref{eq:T21pert} should be multiplied by a common factor of~$e^{-\tau_\mathrm{reio}}$, where~$\tau_\mathrm{reio}$ is the reionization optical depth.
Therefore, the angular power spectra of 21-cm brightness temperature is expected to be damped by a factor~$e^{-2\tau_\mathrm{reio}} \sim 0.9$, approximately, according to the latest Planck data release~\cite{aghanim:planckcosmoparameters}.
As for the effect of finite width of the 21-cm line, we expect that not including this effect will have a minimal impact on our results, especially in the case where the level of the noise is not known to a percent-level accuracy.
If included, this effect would realistically increase the error bars by a few percent of their current values reported in appendix~\ref{app:tables_of_constraints}, preserving the overall conclusions.
Moreover, we expect this additional exponential damping of the signal to be extremely degenerate with the amplitude of primordial fluctuations, making it effectively equivalent to a constant rescaling of the latter.


\section{Angular resolution in presence of foregrounds}
\label{app:resolution_with_foregrounds}

Radio interferometers observe modes within a cylindrical volume in Fourier space given by
\begin{equation}
    V_k \simeq \pi \left( k^2_{\perp,\mathrm{max}} - k^2_{\perp,\mathrm{min}} \right)\left( k_{\parallel,\mathrm{max}} - k_{\parallel,\mathrm{min}}\right),
\end{equation}
where~$k_{\perp,\mathrm{max}}$ and~$k_{\perp,\mathrm{min}}$ are determined by the field of view and the array of antennas, respectively, while~$k_{\parallel,\mathrm{max}}$ and~$k_{\parallel,\mathrm{min}}$ depend on the spectral resolution of the instrument and the bandwidth chosen for the analysis.
In a full-sky survey we can safely approximate~$k_{\perp,\mathrm{min}} \simeq 0$, thus we neglect it in the following derivation.
The volume occupied by the foreground wedge in each frequency channel, i.e., the region satisfying equation~\eqref{eq:foreground_wedge}, is
\begin{equation}
    V_\mathrm{wedge} \simeq \frac{2}{3} \pi k^2_{\perp,\mathrm{max}} \left( k_\mathrm{wedge,max} - k_{\parallel,\mathrm{min}}\right),
\end{equation}
where
\begin{equation}
    k_\mathrm{wedge,max} = \theta_0 d_A(z_i) H(z_i) k_{\perp,\mathrm{max}},
\end{equation}
while the foreground-free volume reads as
\begin{equation}
    V_\mathrm{ff} = V_k - V_\mathrm{wedge} \simeq \pi k^2_{\perp,\mathrm{max}} \left[ k_{\parallel,\mathrm{max}} - k_\mathrm{wedge,max} + \frac{1}{3}\left( k_\mathrm{wedge,max} - k_{\parallel,\mathrm{min}} \right) \right].
\end{equation}
We define a maximum resolution~$\ell_{i,\mathrm{for}}$ to minimize the impact of Fourier modes contaminated by foregrounds in the estimation of the angular power spectra.
Specifically, we require that the volume of foreground-free modes dominates over the volume of the wedge, i.e., that~$V_\mathrm{ff} \gtrsim \alpha_V V_\mathrm{wedge}$, where~$\alpha_V$ is an arbitrary numerical coefficient. 
This requirement explicitly reads as
\begin{equation}
    k_{\parallel,\mathrm{max}} \gtrsim \frac{2\theta_0}{3} (\alpha_V+1) d_A(z_i) H(z_i) k_{\perp,\mathrm{max}},
\end{equation}
or, in terms of multipole resolution, as
\begin{equation}
    \ell_{i,\mathrm{for}} \lesssim \frac{3 \pi}{\theta_0 (\alpha_V+1)} \frac{\nu_i}{\delta\nu},
\end{equation}
where we use~$\ell_{i,\mathrm{for}} \simeq k_{\perp,\mathrm{max}}r_i$ and~$k_{\parallel,\mathrm{max}} \simeq 2\pi \frac{H(z_i) \nu_i}{(1+z_i) \delta\nu}$.
In this work we assume the conservative value of~$\alpha_V = 20$, i.e., only less than~$5\%$ of $k$ modes are impacted by foregrounds, recovering equation~\eqref{eq:maximum_foreground_multipole}.


\section{Tables of constraints and cumulative errors}
\label{app:tables_of_constraints}

\begin{figure}[ht]
    \centerline{
    \includegraphics[width=\columnwidth]{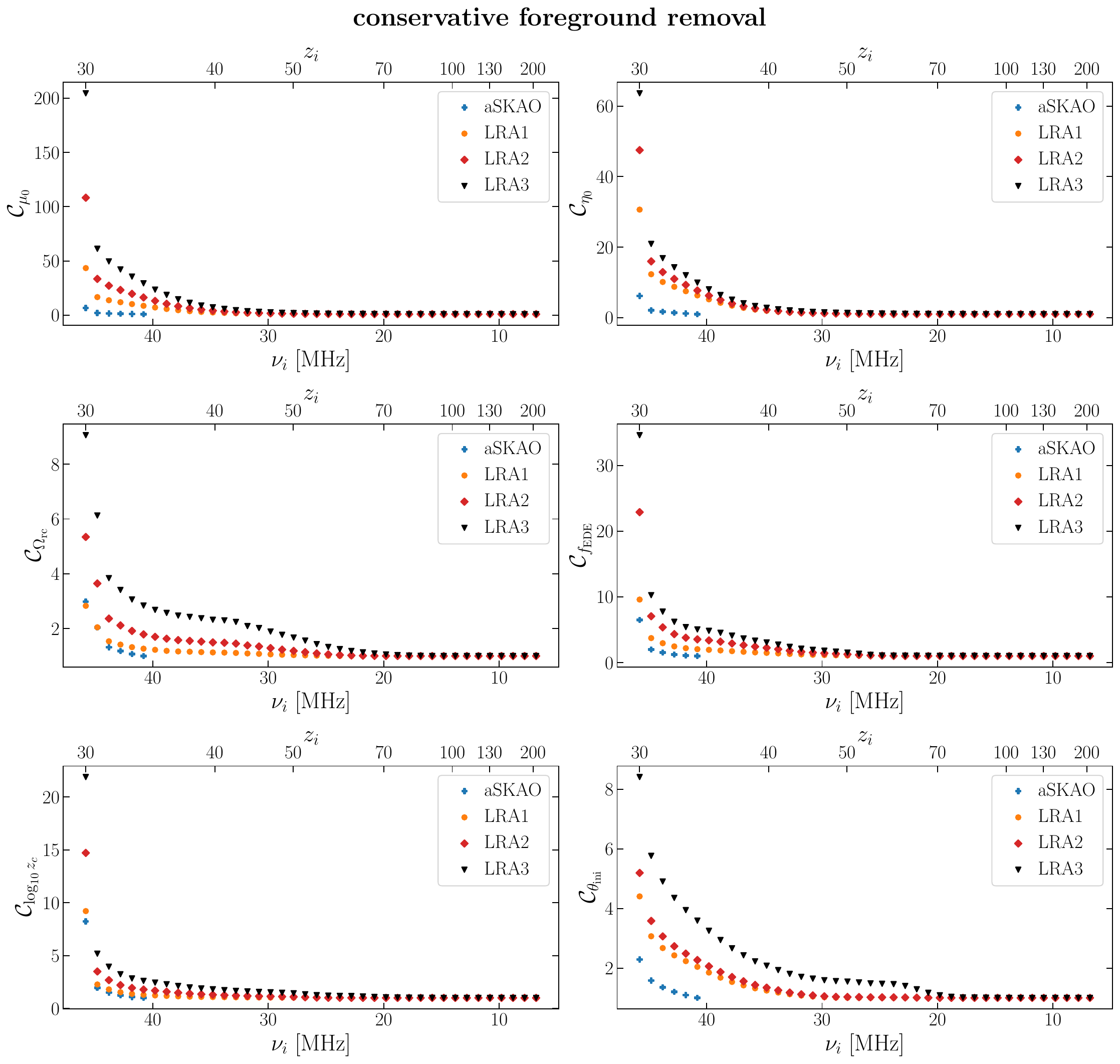}}
    \caption{Cumulative error ratio of equation~\eqref{eq:cumulative_error_ratio} obtained by including only the first~$N$ frequency bins with respect to the~$N_\nu$ total number of frequency bins for the parameters of~$\mu-\eta$, nDGP and EDE models, employing the conservative foreground removal strategy, see equation~\eqref{eq:maximum_foreground_multipole} and the discussion therein. 
    Regarding the~$\mu-\eta$ parametrization, we show only the~$k_*=10^{-1}$ Mpc$^{-1}$ case since we obtain similar results for all~$k_*$ values.}
    \label{fig:cumulative errors_cons}
\end{figure}

\begin{figure}[ht]
    \centerline{
    \includegraphics[width=\columnwidth]{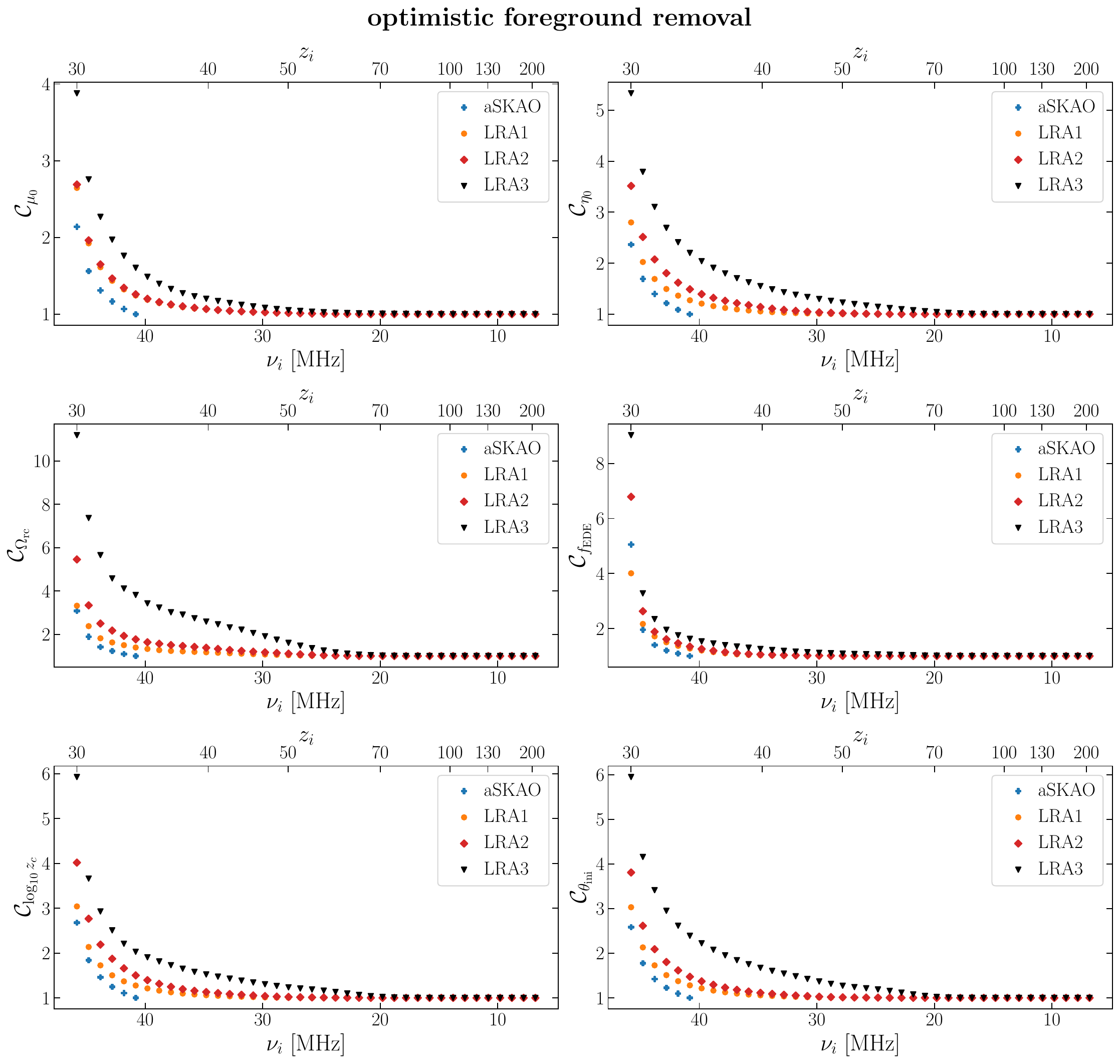}}
    \caption{Same as figure~\ref{fig:cumulative errors_cons}, but for the optimistic foreground removal strategy, see equation~\eqref{eq:maximum_foreground_multipole}.}
    \label{fig:cumulative errors_opt}
\end{figure}

In this appendix we report the numerical values of the constraints presented in the figures of section~\ref{sec:constraints_growth}.
Moreover, to highlight the constraining power obtained by including additional frequency channels to a given experiment, we show the cumulative error ratio defined in equation~\eqref{eq:cumulative_error_ratio} in figures~\ref{fig:cumulative errors_cons} and~\ref{fig:cumulative errors_opt}, for the conservative and optimistic foreground removal cases respectively.

\begin{table}[ht]
\centerline{
\begin{tabular}{|c|c|cc|cc|}
    \hline
    \multirow{2}{*}{Experiment}  & \multirow{2}{*}{$k_*\ [\mathrm{Mpc}^{-1}]$} &  \multicolumn{2}{|c|}{conservative} & \multicolumn{2}{|c|}{optimistic}    \\
           &  & $\sigma_{\mu_0}$ & $\sigma_{\eta_0}$ & $\sigma_{\mu_0}$ & $\sigma_{\eta_0}$ \\ 
    \hline
    \hline 
    \multirow{4}{*}{aSKAO} & $\to \infty$ & $4.21\times10^{-1}$  & $8.41\times10^{-1}$ 
                                          & $1.68\times10^{-2}$  & $5.02\times10^{-2}$  \\
                           & $1$       & $4.21\times10^{-1}$  &$8.49\times10^{-1}$   
                                       & $1.23\times10^{-2}$  & $5.33\times10^{-2}$    \\
                           & $10^{-1}$ & $3.08\times10^{-1}$  &$1.09$  
                                       & $2.41\times10^{-3}$  & $5.45\times10^{-2}$  \\
                           & $10^{-2}$ & $9.75\times10^{-2}$  &$2.20\times10^{-1}$  
                                        & $2.53\times10^{-2}$  & $1.74\times10^{-1}$  \\ 
    \hline
    \multirow{4}{*}{LRA1} & $\to \infty$ &$2.83\times10^{-1}$  & $1.04$  
                                         & $1.45\times10^{-1}$  & $4.60\times10^{-1}$ \\
                          & $1$       &$2.83\times10^{-1}$ &  $1.04$    
                                      & $1.46\times10^{-1}$  & $4.28\times10^{-1}$ \\
                          & $10^{-1}$ &$2.95\times10^{-1}$  & $1.38$     
                                        & $5.14\times10^{-2}$  & $5.40\times10^{-1}$ \\
                          & $10^{-2}$ &$4.78\times10^{-1}$  & $8.13\times10^{-1}$
                                       & $1.49\times10^{-1}$  & $6.94\times10^{-1}$ \\ 
    \hline
    \multirow{4}{*}{LRA2} & $\to \infty$ &$1.78\times10^{-2}$  &$1.21\times10^{-1}$   
                                         & $5.39\times10^{-3}$  & $3.19\times10^{-2}$ \\
                          & $1$        &$1.18\times10^{-2}$  & $1.21\times10^{-1}$  
                                       & $5.01\times10^{-3}$  & $3.27\times10^{-2}$  \\
                          & $10^{-1}$  &$1.88\times10^{-2}$  &$1.41\times10^{-1}$  
                                       & $1.92\times10^{-3}$  & $3.66\times10^{-2}$  \\
                          & $10^{-2}$  &$5.39\times10^{-2}$  &$1.10\times10^{-1}$  
                                       & $1.60\times10^{-2}$  & $9.67\times10^{-2}$  \\ 
    \hline
    \multirow{4}{*}{LRA3} & $\to \infty$ &$8.12\times10^{-3}$  & $8.10\times10^{-2}$  
                                         & $3.89\times10^{-4}$  & $1.38\times10^{-2}$ \\
                          & $1$       &$8.13\times10^{-3}$ & $8.11\times10^{-2}$ 
                                      & $3.51\times10^{-4}$  & $1.45\times10^{-2}$  \\
                          & $10^{-1}$ &$8.57\times10^{-3}$  & $9.03\times10^{-2}$  
                                      & $2.79\times10^{-4}$  & $1.31\times10^{-2}$  \\
                          & $10^{-2}$ &$3.30\times10^{-2}$  & $6.44\times10^{-2}$ 
                                      & $9.89\times10^{-3}$  & $5.97\times10^{-2}$ \\ 
    \hline
\end{tabular}}
\caption{Marginalized errors on~$\mu_0$ and~$\eta_0$ for aSKAO, LRA1, LRA2 and LRA3 in the conservative and optimistic case.
For a visual comparison, we refer the reader to the central and right panel of figure~\ref{fig:mPk_errors_mueta}.Fiducial values of the parameters used for the Fisher forecast:~$\{\mu_0=1, \eta_0=1\}$;~$\Lambda$CDM parameters as listed in equation~\eqref{eq:lcdm_fiducial_reconstruction}}
\label{tab:mu0eta0_constraints}
\end{table}

\begin{table}[ht]
\centerline{
\begin{tabular}{|c|c|c|}
    \hline
    \multirow{2}{*}{Experiment}  & conservative & optimistic   \\
                 & $\sigma_{\Omega_\mathrm{rc}}$ & $\sigma_{\Omega_\mathrm{rc}}$   \\
    \hline
    \hline
    aSKAO& $6.11$ &  $2.18$ \\
    LRA1 & $37.32$ &    $26.13$  \\
    LRA2 & $3.42$ &   $1.23$  \\ 
    LRA3 & $1.79\times10^{-1}$ &  $1.25\times10^{-1}$  \\
    \hline
\end{tabular}}
\caption{Marginalized errors on $\Omega_\mathrm{rc}$ for aSKAO, LRA1, LRA2 and LRA3 in the conservative and optimistic case.
For a visual comparison, we refer the reader to the central and right panels of figure~\ref{fig:mPk_errors_nDGP}. Fiducial values of the parameters used for the Fisher forecast:~$\{\Omega_{\rm{rc}}=0.2\}$;~$\Lambda$CDM parameters as listed in equation~\eqref{eq:lcdm_fiducial_reconstruction}.}
\label{tab:nDGP_constraints}
\end{table}

\begin{table}[ht]
\centerline{
\begin{tabular}{|c|ccc|ccc|}
    \hline
    \multirow{2}{*}{Experiment}  & \multicolumn{3}{|c|}{conservative} & \multicolumn{3}{|c|}{optimistic}    \\
           & $\sigma_{f_\mathrm{EDE}}$ & $\sigma_{\log_{10} z_c}$ & $\sigma_{\theta_\mathrm{ini}}$ & $\sigma_{f_\mathrm{EDE}}$ & $\sigma_{\log_{10} z_c}$ & $\sigma_{\theta_\mathrm{ini}}$  \\
    \hline
    \hline
    aSKAO& $1.02$  &  $1.41$  & $5.91$ & $7.21\times10^{-3}$  &  $1.18\times10^{-1}$  & $6.26\times10^{-1}$\\
    LRA1 & $3.74$  & $7.09$  & $26.14$  & $1.35\times10^{-1}$  & $1.75$  & $8.65$  \\
    LRA2 &  $2.88\times10^{-1}$  & $7.90\times10^{-1}$  & $4.60$&  $5.36\times10^{-3}$  & $7.86\times10^{-2}$  & $4.25\times10^{-1}$  \\ 
    LRA3 & $1.70\times10^{-1}$  & $4.72\times10^{-1}$  &  $1.34$ & $7.98\times10^{-4}$  & $1.18\times10^{-2}$  &  $1.02\times10^{-1}$ \\
    \hline
\end{tabular}}
\caption{Marginalized errors on $f_{\rm EDE}$, $\log_{10} z_c$ and~$\theta_\mathrm{ini}$ for aSKAO, LRA1, LRA2 and LRA3 in the conservative and optimistic case.
For a visual comparison, we refer the reader to the central and right panels of figure~\ref{fig:mPk_errors_EDE}. Fiducial values of the parameters used for the Fisher forecast:$\{ f_{\rm EDE} = 0.122, \log_{10} z_c = 3.562, \theta_\mathrm{ini} = 2.83 \}$;~$\Lambda$CDM parameters as listed in section~\ref{subsec:ede_theory}.
}
\label{tab:EDE_constraints}
\end{table}

\bibliography{bibliography}

\providecommand{\href}[2]{#2}\begingroup\raggedright\begin{thebibliography}{100}

\bibitem{peebles:lssformation}
P.~J.~E. {Peebles}, ``Large-scale background temperature and mass fluctuations
  due to scale-invariant primeval perturbations'',
  \href{http://dx.doi.org/10.1086/183911}{{\em The Astrophysical Journal
  Letters} {\bfseries 263} (Dec, 1982) L1--L5}.

\bibitem{blumenthal:lssformation}
G.~R. Blumenthal, S.~M. Faber, J.~R. Primack, and M.~J. Rees, ``Formation of
  galaxies and large-scale structure with cold dark matter'',
  \href{http://dx.doi.org/0.1038/311517a0}{{\em Nature} {\bfseries 311} (Oct,
  1984) 517--525}.

\bibitem{davis:lssformation}
M.~{Davis}, G.~{Efstathiou}, C.~S. {Frenk}, and S.~D.~M. {White}, ``The
  evolution of large-scale structure in a universe dominated by cold dark
  matter'', \href{http://dx.doi.org/10.1086/163168}{{\em The Astrophysical
  Journal} {\bfseries 292} (May, 1985) 371--394}.

\bibitem{bernardeau:lssreview}
F.~Bernardeau, S.~Colombi, E.~Gaztañaga, and R.~Scoccimarro, ``Large-scale
  structure of the Universe and cosmological perturbation theory'',
  \href{http://dx.doi.org/10.1016/S0370-1573(02)00135-7}{{\em Physics Reports}
  {\bfseries 367} no.~1, (2002) 1--248},
  \href{http://arxiv.org/abs/astro-ph/0112551}{{\ttfamily
  arXiv:astro-ph/0112551}}.

\bibitem{cooray:halomodelreview}
A.~Cooray and R.~Sheth, ``Halo models of large scale structure'',
  \href{http://dx.doi.org/10.1016/S0370-1573(02)00276-4}{{\em Physics Reports}
  {\bfseries 372} no.~1, (2002) 1--129},
  \href{http://arxiv.org/abs/astro-ph/0206508}{{\ttfamily
  arXiv:astro-ph/0206508}}.

\bibitem{aghanim:planckcosmoparameters}
The {\bfseries Planck Collaboration}, N.~Aghanim {\em et~al.}, ``Planck 2018
  results - VI. Cosmological parameters'',
  \href{http://dx.doi.org/10.1051/0004-6361/201833910}{{\em A\&A} {\bfseries
  641} (2020) A6}, \href{http://arxiv.org/abs/1807.06209}{{\ttfamily
  arXiv:1807.06209}}.

\bibitem{aiola:actcosmoparameters}
S.~Aiola {\em et~al.}, ``The Atacama Cosmology Telescope: DR4 maps and
  cosmological parameters'',
  \href{http://dx.doi.org/10.1088/1475-7516/2020/12/047}{{\em Journal of
  Cosmology and Astroparticle Physics} {\bfseries 2020} no.~12, (Dec, 2020)
  047}, \href{http://arxiv.org/abs/2007.07288}{{\ttfamily arXiv:2007.07288}}.

\bibitem{pan:sptcosmoparameters}
Z.~Pan {\em et~al.}, ``Measurement of gravitational lensing of the cosmic
  microwave background using SPT-3G 2018 data'',
  \href{http://dx.doi.org/10.1103/PhysRevD.108.122005}{{\em Phys. Rev. D}
  {\bfseries 108} (Dec, 2023) 122005},
  \href{http://arxiv.org/abs/2308.11608}{{\ttfamily arXiv:2308.11608}}.

\bibitem{hawkins:2dfsurvey}
E.~Hawkins {\em et~al.}, ``The 2dF Galaxy Redshift Survey: correlation
  functions, peculiar velocities and the matter density of the Universe'',
  \href{http://dx.doi.org/10.1046/j.1365-2966.2003.07063.x}{{\em Monthly
  Notices of the Royal Astronomical Society} {\bfseries 346} no.~1, (11, 2003)
  78--96}, \href{http://arxiv.org/abs/astro-ph/0212375}{{\ttfamily
  arXiv:astro-ph/0212375}}.

\bibitem{tegmark:sdsssurvey}
M.~Tegmark {\em et~al.}, ``Cosmological constraints from the SDSS luminous red
  galaxies'', \href{http://dx.doi.org/10.1103/PhysRevD.74.123507}{{\em Phys.
  Rev. D} {\bfseries 74} (Dec, 2006) 123507},
  \href{http://arxiv.org/abs/astro-ph/0608632}{{\ttfamily
  arXiv:astro-ph/0608632}}.

\bibitem{ross:2dfsdsssurvey}
N.~P. Ross {\em et~al.}, ``The 2dF-SDSS LRG and QSO Survey: the LRG 2-point
  correlation function and redshift-space distortions'',
  \href{http://dx.doi.org/10.1111/j.1365-2966.2007.12289.x}{{\em Monthly
  Notices of the Royal Astronomical Society} {\bfseries 381} no.~2, (10, 2007)
  573--588}, \href{http://arxiv.org/abs/astro-ph/0612400}{{\ttfamily
  arXiv:astro-ph/0612400}}.

\bibitem{daangela:2dfsdsssurvey}
J.~da~Angela {\em et~al.}, ``The 2dF-SDSS LRG and QSO survey: QSO clustering
  and the L–z degeneracy'',
  \href{http://dx.doi.org/10.1111/j.1365-2966.2007.12552.x}{{\em Monthly
  Notices of the Royal Astronomical Society} {\bfseries 383} no.~2, (12, 2007)
  565--580}, \href{http://arxiv.org/abs/astro-ph/0612401}{{\ttfamily
  arXiv:astro-ph/0612401}}.

\bibitem{guzzo:2dfvimossurvey}
L.~Guzzo {\em et~al.}, ``A test of the nature of cosmic acceleration using
  galaxy redshift distortions'',
  \href{http://dx.doi.org/10.1038/nature06555}{{\em Nature} {\bfseries 451}
  (Jan, 2008) 541--544}, \href{http://arxiv.org/abs/0802.1944}{{\ttfamily
  arXiv:0802.1944}}.

\bibitem{blake:wigglezsurvey}
C.~Blake {\em et~al.}, ``The WiggleZ Dark Energy Survey: the growth rate of
  cosmic structure since redshift z=0.9'',
  \href{http://dx.doi.org/10.1111/j.1365-2966.2011.18903.x}{{\em Monthly
  Notices of the Royal Astronomical Society} {\bfseries 415} no.~3, (08, 2011)
  2876--2891}, \href{http://arxiv.org/abs/1104.2948}{{\ttfamily
  arXiv:1104.2948}}.

\bibitem{blake:gamasurvey}
C.~Blake {\em et~al.}, ``Galaxy And Mass Assembly (GAMA): improved cosmic
  growth measurements using multiple tracers of large-scale structure'',
  \href{http://dx.doi.org/10.1093/mnras/stt1791}{{\em Monthly Notices of the
  Royal Astronomical Society} {\bfseries 436} no.~4, (10, 2013) 3089--3105},
  \href{http://arxiv.org/abs/1309.5556}{{\ttfamily arXiv:1309.5556}}.

\bibitem{blake:wigglezsurveyII}
C.~Blake {\em et~al.}, ``The WiggleZ Dark Energy Survey: joint measurements of
  the expansion and growth history at z \&lt; 1'',
  \href{http://dx.doi.org/10.1111/j.1365-2966.2012.21473.x}{{\em Monthly
  Notices of the Royal Astronomical Society} {\bfseries 425} no.~1, (09, 2012)
  405--414}, \href{http://arxiv.org/abs/1204.3674}{{\ttfamily
  arXiv:1204.3674}}.

\bibitem{beutler:6dfsurvey}
F.~Beutler {\em et~al.}, ``The 6dF Galaxy Survey: $z\approx 0$ measurements of
  the growth rate and $\sigma_8$'',
  \href{http://dx.doi.org/10.1111/j.1365-2966.2012.21136.x}{{\em Monthly
  Notices of the Royal Astronomical Society} {\bfseries 423} no.~4, (07, 2012)
  3430--3444}, \href{http://arxiv.org/abs/1204.4725}{{\ttfamily
  arXiv:1204.4725}}.

\bibitem{sanchez:sdssiiisurvey}
A.~G. S\'anchez {\em et~al.}, ``The clustering of galaxies in the SDSS-III
  Baryon Oscillation Spectroscopic Survey: cosmological implications of the
  full shape of the clustering wedges in the data release 10 and 11 galaxy
  samples'', \href{http://dx.doi.org/10.1093/mnras/stu342}{{\em Monthly Notices
  of the Royal Astronomical Society} {\bfseries 440} no.~3, (04, 2014)
  2692--2713}, \href{http://arxiv.org/abs/1312.4854}{{\ttfamily
  arXiv:1312.4854}}.

\bibitem{chuang:sdssiiisurvey}
C.-H. Chuang {\em et~al.}, ``The clustering of galaxies in the SDSS-III Baryon
  Oscillation Spectroscopic Survey: single-probe measurements from CMASS
  anisotropic galaxy clustering'',
  \href{http://dx.doi.org/10.1093/mnras/stw1535}{{\em Monthly Notices of the
  Royal Astronomical Society} {\bfseries 461} no.~4, (06, 2016) 3781--3793},
  \href{http://arxiv.org/abs/1312.4889}{{\ttfamily arXiv:1312.4889}}.

\bibitem{feix:sdsssurvey}
M.~Feix, A.~Nusser, and E.~Branchini, ``Growth Rate of Cosmological
  Perturbations at $z\ensuremath{\sim}0.1$ from a New Observational Test'',
  \href{http://dx.doi.org/10.1103/PhysRevLett.115.011301}{{\em Phys. Rev.
  Lett.} {\bfseries 115} (Jun, 2015) 011301},
  \href{http://arxiv.org/abs/1503.05945}{{\ttfamily arXiv:1503.05945}}.

\bibitem{okumura:fmossurvey}
T.~Okumura {\em et~al.}, ``The Subaru FMOS galaxy redshift survey (FastSound).
  IV. New constraint on gravity theory from redshift space distortions at $z
  \sim 1.4$'', \href{http://dx.doi.org/10.1093/pasj/psw029}{{\em Publications
  of the Astronomical Society of Japan} {\bfseries 68} no.~3, (04, 2016) 38},
  \href{http://arxiv.org/abs/1511.08083}{{\ttfamily arXiv:1511.08083}}.

\bibitem{achitouv:6dfsurvey}
I.~Achitouv, C.~Blake, P.~Carter, J.~Koda, and F.~Beutler, ``Consistency of the
  growth rate in different environments with the 6-degree Field Galaxy Survey:
  Measurement of the void-galaxy and galaxy-galaxy correlation functions'',
  \href{http://dx.doi.org/10.1103/PhysRevD.95.083502}{{\em Phys. Rev. D}
  {\bfseries 95} (Apr, 2017) 083502},
  \href{http://arxiv.org/abs/1606.03092}{{\ttfamily arXiv:1606.03092}}.

\bibitem{alam:sdssiiisurvey}
S.~Alam {\em et~al.}, ``The clustering of galaxies in the completed SDSS-III
  Baryon Oscillation Spectroscopic Survey: cosmological analysis of the DR12
  galaxy sample'', \href{http://dx.doi.org/10.1093/mnras/stx721}{{\em Monthly
  Notices of the Royal Astronomical Society} {\bfseries 470} no.~3, (03, 2017)
  2617--2652}, \href{http://arxiv.org/abs/1607.03155}{{\ttfamily
  arXiv:1607.03155}}.

\bibitem{zhao:sdssivsurvey}
G.-B. Zhao {\em et~al.}, ``The clustering of the SDSS-IV extended Baryon
  Oscillation Spectroscopic Survey DR14 quasar sample: a tomographic
  measurement of cosmic structure growth and expansion rate based on optimal
  redshift weights'', \href{http://dx.doi.org/10.1093/mnras/sty2845}{{\em
  Monthly Notices of the Royal Astronomical Society} {\bfseries 482} no.~3,
  (10, 2018) 3497--3513}, \href{http://arxiv.org/abs/1801.03043}{{\ttfamily
  arXiv:1801.03043}}.

\bibitem{nadathur:bosssurvey}
S.~Nadathur, P.~M. Carter, W.~J. Percival, H.~A. Winther, and J.~E. Bautista,
  ``Beyond BAO: Improving cosmological constraints from BOSS data with
  measurement of the void-galaxy cross-correlation'',
  \href{http://dx.doi.org/10.1103/PhysRevD.100.023504}{{\em Phys. Rev. D}
  {\bfseries 100} (Jul, 2019) 023504},
  \href{http://arxiv.org/abs/1904.01030}{{\ttfamily arXiv:1904.01030}}.

\bibitem{aubert:sdssivsurvey}
M.~Aubert {\em et~al.}, ``The completed SDSS-IV extended Baryon Oscillation
  Spectroscopic Survey: growth rate of structure measurement from cosmic
  voids'', \href{http://dx.doi.org/10.1093/mnras/stac828}{{\em Monthly Notices
  of the Royal Astronomical Society} {\bfseries 513} no.~1, (03, 2022)
  186--203}, \href{http://arxiv.org/abs/2007.09013}{{\ttfamily
  arXiv:2007.09013}}.

\bibitem{Samushia:2012}
L.~{Samushia}, W.~J. {Percival}, and A.~{Raccanelli}, ``Interpreting
  large-scale redshift-space distortion measurements'',
  \href{http://dx.doi.org/10.1111/j.1365-2966.2011.20169.x}{{\em Monthly
  Notices of the Royal Astronomical Society} {\bfseries 420} no.~3, (Mar, 2012)
  2102--2119}, \href{http://arxiv.org/abs/1102.1014}{{\ttfamily
  arXiv:1102.1014}}.

\bibitem{Raccanelli:2013}
A.~{Raccanelli}, D.~{Bertacca}, D.~{Pietrobon}, F.~{Schmidt}, L.~{Samushia},
  N.~{Bartolo}, O.~{Dor{\'e}}, S.~{Matarrese}, and W.~J. {Percival}, ``{Testing
  gravity using large-scale redshift-space distortions}'',
  \href{http://dx.doi.org/10.1093/mnras/stt1517}{{\em Monthly Notices of the
  Royal Astronomical Society} {\bfseries 436} no.~1, (Nov, 2013) 89--100},
  \href{http://arxiv.org/abs/1207.0500}{{\ttfamily arXiv:1207.0500}}.

\bibitem{Huterer:2015}
D.~{Huterer} {\em et~al.}, ``Growth of cosmic structure: Probing dark energy
  beyond expansion'',
  \href{http://dx.doi.org/10.1016/j.astropartphys.2014.07.004}{{\em
  Astroparticle Physics} {\bfseries 63} (Mar, 2015) 23--41},
  \href{http://arxiv.org/abs/1309.5385}{{\ttfamily arXiv:1309.5385}}.

\bibitem{DESI:2024hhd}
The {\bfseries DESI Collaboration}, A.~G. Adame {\em et~al.}, ``{DESI 2024 VII:
  cosmological constraints from the full-shape modeling of clustering
  measurements}'', \href{http://dx.doi.org/10.1088/1475-7516/2025/07/028}{{\em
  JCAP} {\bfseries 07} (2025) 028},
  \href{http://arxiv.org/abs/2411.12022}{{\ttfamily arXiv:2411.12022
  [astro-ph.CO]}}.

\bibitem{DESI:2024mwx}
The {\bfseries DESI Collaboration}, A.~G. Adame {\em et~al.}, ``{DESI 2024 VI:
  cosmological constraints from the measurements of baryon acoustic
  oscillations}'', \href{http://dx.doi.org/10.1088/1475-7516/2025/02/021}{{\em
  JCAP} {\bfseries 02} (2025) 021},
  \href{http://arxiv.org/abs/2404.03002}{{\ttfamily arXiv:2404.03002
  [astro-ph.CO]}}.

\bibitem{heymans:sigma8tension}
C.~Heymans {\em et~al.}, ``KiDS-1000 Cosmology: Multi-probe weak gravitational
  lensing and spectroscopic galaxy clustering constraints'',
  \href{http://dx.doi.org/10.1051/0004-6361/202039063}{{\em A\&A} {\bfseries
  646} (2021) A140}, \href{http://arxiv.org/abs/2007.15632}{{\ttfamily
  arXiv:2007.15632}}.

\bibitem{abbott:sigma8tension}
The {\bfseries DES Collaboration}, T.~M.~C. Abbott {\em et~al.}, ``Dark Energy
  Survey Year 3 results: Cosmological constraints from galaxy clustering and
  weak lensing'', \href{http://dx.doi.org/10.1103/PhysRevD.105.023520}{{\em
  Phys. Rev. D} {\bfseries 105} (Jan, 2022) 023520},
  \href{http://arxiv.org/abs/2105.13549}{{\ttfamily arXiv:2105.13549}}.

\bibitem{verde:H0review}
L.~Verde, N.~Sch\"oneberg, and H.~Gil-Mar\'in, ``A Tale of Many $H_0$'',
  \href{http://dx.doi.org/10.1146/annurev-astro-052622-033813}{{\em Annual
  Review of Astronomy and Astrophysics} {\bfseries 62} (2024) 287--331},
  \href{http://arxiv.org/abs/2311.13305}{{\ttfamily arXiv:2311.13305}}.

\bibitem{divalentino:H0solutions}
E.~Di~Valentino, O.~Mena, S.~Pan, L.~Visinelli, W.~Yang, A.~Melchiorri, D.~F.
  Mota, A.~G. Riess, and J.~Silk, ``In the realm of the Hubble
  tension\textemdash{}a review of solutions'',
  \href{http://dx.doi.org/10.1088/1361-6382/ac086d}{{\em Class. Quant. Grav.}
  {\bfseries 38} no.~15, (2021) 153001},
  \href{http://arxiv.org/abs/2103.01183}{{\ttfamily arXiv:2103.01183}}.

\bibitem{schonenberg:H0olympics}
N.~Sch\"oneberg, G.~F. Abell\'an, A.~P. S\'anchez, S.~J. Witte, V.~Poulin, and
  J.~Lesgourgues, ``The $H_0$ Olympics: A fair ranking of proposed models'',
  \href{http://dx.doi.org/10.1016/j.physrep.2022.07.001}{{\em Physics Reports}
  {\bfseries 984} (2022) 1--55},
  \href{http://arxiv.org/abs/2107.10291}{{\ttfamily arXiv:2107.10291}}.

\bibitem{khalife:H0solution}
A.~R. Khalife, M.~B. Zanjani, S.~Galli, S.~G\"unther, J.~Lesgourgues, and
  K.~Benabed, ``Review of Hubble tension solutions with new SH0ES and SPT-3G
  data'', \href{http://dx.doi.org/10.1088/1475-7516/2024/04/059}{{\em JCAP}
  {\bfseries 04} (2024) 059}, \href{http://arxiv.org/abs/2312.09814}{{\ttfamily
  arXiv:2312.09814}}.

\bibitem{kovetz:limstatusreport}
E.~D. Kovetz {\em et~al.}, ``Line-Intensity Mapping: 2017 Status Report'',
  \href{http://arxiv.org/abs/1709.09066}{{\ttfamily arXiv:1709.09066}}.

\bibitem{kovetz:astrocosmolim}
E.~D. Kovetz {\em et~al.}, ``Astrophysics and Cosmology with Line-Intensity
  Mapping'', {\em Bull. Am. Astron. Soc.} {\bfseries 51} no.~3, (2020) 101,
  \href{http://arxiv.org/abs/1903.04496}{{\ttfamily arXiv:1903.04496}}.

\bibitem{silva:limlss}
M.~B. Silva, E.~D. Kovetz, G.~K. Keating, A.~Moradinezhad~Dizgah, M.~Bethermin,
  P.~C. Breysse, K.~Kartare, J.~L. Bernal, and J.~Delabrouille, ``Mapping
  large-scale-structure evolution over cosmic times'',
  \href{http://dx.doi.org/10.1007/s10686-021-09755-3}{{\em Exper. Astron.}
  {\bfseries 51} no.~3, (2021) 1593--1622},
  \href{http://arxiv.org/abs/1908.07533}{{\ttfamily arXiv:1908.07533}}.

\bibitem{Masui:2009cj}
K.~W. Masui, F.~Schmidt, U.-L. Pen, and P.~McDonald, ``{Projected Constraints
  on Modified Gravity Cosmologies from 21cm Intensity Mapping}'',
  \href{http://dx.doi.org/10.1103/PhysRevD.81.062001}{{\em Phys. Rev. D}
  {\bfseries 81} (2010) 062001},
  \href{http://arxiv.org/abs/0911.3552}{{\ttfamily arXiv:0911.3552}}.

\bibitem{Bull:2014rha}
P.~Bull, P.~G. Ferreira, P.~Patel, and M.~G. Santos, ``{Late-time cosmology
  with 21cm intensity mapping experiments}'',
  \href{http://dx.doi.org/10.1088/0004-637X/803/1/21}{{\em Astrophys. J.}
  {\bfseries 803} no.~1, (2015) 21},
  \href{http://arxiv.org/abs/1405.1452}{{\ttfamily arXiv:1405.1452}}.

\bibitem{Karagiannis:2022ylq}
D.~Karagiannis, R.~Maartens, and L.~F. Randrianjanahary, ``{Cosmological
  constraints from the power spectrum and bispectrum of 21cm intensity maps}'',
  \href{http://dx.doi.org/10.1088/1475-7516/2022/11/003}{{\em JCAP} {\bfseries
  11} (2022) 003}, \href{http://arxiv.org/abs/2206.07747}{{\ttfamily
  arXiv:2206.07747}}.

\bibitem{Scott:2022fev}
B.~R. Scott, K.~S. Karkare, and S.~Bird, ``{A forecast for large-scale
  structure constraints on Horndeski gravity with CO line intensity mapping}'',
  \href{http://dx.doi.org/10.1093/mnras/stad1501}{{\em Mon. Not. Roy. Astron.
  Soc.} {\bfseries 523} no.~4, (2023) 4895--4908},
  \href{http://arxiv.org/abs/2209.13029}{{\ttfamily arXiv:2209.13029}}.

\bibitem{Casas:2022vik}
S.~Casas, I.~P. Carucci, V.~Pettorino, S.~Camera, and M.~Martinelli,
  ``{Constraining gravity with synergies between radio and optical cosmological
  surveys}'', \href{http://dx.doi.org/10.1016/j.dark.2022.101151}{{\em Phys.
  Dark Univ.} {\bfseries 39} (2023) 101151},
  \href{http://arxiv.org/abs/2210.05705}{{\ttfamily arXiv:2210.05705}}.

\bibitem{Castorina:2019zho}
E.~Castorina and M.~White, ``{Measuring the growth of structure with intensity
  mapping surveys}'',
  \href{http://dx.doi.org/10.1088/1475-7516/2019/06/025}{{\em JCAP} {\bfseries
  06} (2019) 025}, \href{http://arxiv.org/abs/1902.07147}{{\ttfamily
  arXiv:1902.07147}}.

\bibitem{MoradinezhadDizgah:2023src}
A.~Moradinezhad~Dizgah, E.~Bellini, and G.~K. Keating, ``{Probing Dark Energy
  and Modifications of Gravity with Ground-based millimeter-wavelength Line
  Intensity Mapping}'', \href{http://dx.doi.org/10.3847/1538-4357/ad2078}{{\em
  Astrophys. J.} {\bfseries 965} no.~1, (2024) 19},
  \href{http://arxiv.org/abs/2304.08471}{{\ttfamily arXiv:2304.08471}}.

\bibitem{field:21cmline}
G.~B. Field, ``{Excitation of the Hydrogen 21-CM Line}'',
  \href{http://dx.doi.org/10.1109/JRPROC.1958.286741}{{\em IEEE Proc.}
  {\bfseries 46} no.~1, (1958) 240--250}.

\bibitem{zaldarriaga:21cmline}
M.~Zaldarriaga, S.~R. Furlanetto, and L.~Hernquist, ``{21 Centimeter
  fluctuations from cosmic gas at high redshifts}'',
  \href{http://dx.doi.org/10.1086/386327}{{\em Astrophys. J.} {\bfseries 608}
  (2004) 622--635}, \href{http://arxiv.org/abs/astro-ph/0311514}{{\ttfamily
  arXiv:astro-ph/0311514}}.

\bibitem{Furlanetto:2019jso}
S.~Furlanetto {\em et~al.}, ``{Astro 2020 Science White Paper: Fundamental
  Cosmology in the Dark Ages with 21-cm Line Fluctuations}'',
  \href{http://arxiv.org/abs/1903.06212}{{\ttfamily arXiv:1903.06212}}.

\bibitem{cole:darkagesmodes}
P.~S. Cole and J.~Silk, ``Small-scale primordial fluctuations in the 21 cm Dark
  Ages signal'', \href{http://dx.doi.org/10.1093/mnras/staa3638}{{\em Monthly
  Notices of the Royal Astronomical Society} {\bfseries 501} no.~2, (11, 2020)
  2627--2634}, \href{http://arxiv.org/abs/1912.02171}{{\ttfamily
  arXiv:1912.02171}}.

\bibitem{Poulin:2023lkg}
V.~Poulin, T.~L. Smith, and T.~Karwal, ``{The Ups and Downs of Early Dark
  Energy solutions to the Hubble tension: A review of models, hints and
  constraints circa 2023}'',
  \href{http://dx.doi.org/10.1016/j.dark.2023.101348}{{\em Phys. Dark Univ.}
  {\bfseries 42} (2023) 101348},
  \href{http://arxiv.org/abs/2302.09032}{{\ttfamily arXiv:2302.09032}}.

\bibitem{Kamionkowski:2022pkx}
M.~Kamionkowski and A.~G. Riess, ``{The Hubble Tension and Early Dark
  Energy}'', \href{http://dx.doi.org/10.1146/annurev-nucl-111422-024107}{{\em
  Ann. Rev. Nucl. Part. Sci.} {\bfseries 73} (2023) 153--180},
  \href{http://arxiv.org/abs/2211.04492}{{\ttfamily arXiv:2211.04492}}.

\bibitem{Dvali:2000hr}
G.~R. Dvali, G.~Gabadadze, and M.~Porrati, ``{4-D gravity on a brane in 5-D
  Minkowski space}'',
  \href{http://dx.doi.org/10.1016/S0370-2693(00)00669-9}{{\em Phys. Lett. B}
  {\bfseries 485} (2000) 208--214},
  \href{http://arxiv.org/abs/hep-th/0005016}{{\ttfamily arXiv:hep-th/0005016}}.

\bibitem{dekruijf:21cmconstraints}
J.~de~Kruijf, E.~Vanzan, K.~K. Boddy, A.~Raccanelli, and N.~Bartolo,
  ``{Searching for blue-tilted power spectra in the dark ages}'',
  \href{http://dx.doi.org/10.1103/PhysRevD.111.063507}{{\em Phys. Rev. D}
  {\bfseries 111} no.~6, (2025) 063507},
  \href{http://arxiv.org/abs/2408.04991}{{\ttfamily arXiv:2408.04991
  [astro-ph.CO]}}.

\bibitem{Furlanetto2006REVIEW}
S.~Furlanetto, S.~P. Oh, and F.~Briggs, ``{Cosmology at Low Frequencies: The 21
  cm Transition and the High-Redshift Universe}'',
  \href{http://dx.doi.org/10.1016/j.physrep.2006.08.002}{{\em Phys. Rept.}
  {\bfseries 433} (2006) 181--301},
  \href{http://arxiv.org/abs/astro-ph/0608032}{{\ttfamily
  arXiv:astro-ph/0608032}}.

\bibitem{Pritchard:2011xb}
J.~R. Pritchard and A.~Loeb, ``{21-cm cosmology}'',
  \href{http://dx.doi.org/10.1088/0034-4885/75/8/086901}{{\em Rept. Prog.
  Phys.} {\bfseries 75} (2012) 086901},
  \href{http://arxiv.org/abs/1109.6012}{{\ttfamily arXiv:1109.6012}}.

\bibitem{Pen:2008fw}
U.-L. Pen, L.~Staveley-Smith, J.~Peterson, and T.-C. Chang, ``{First Detection
  of Cosmic Structure in the 21-cm Intensity Field}'',
  \href{http://dx.doi.org/10.1111/j.1745-3933.2008.00581.x}{{\em Mon. Not. Roy.
  Astron. Soc.} {\bfseries 394} (2009) 6},
  \href{http://arxiv.org/abs/0802.3239}{{\ttfamily arXiv:0802.3239}}.

\bibitem{Chang:2010jp}
T.-C. Chang, U.-L. Pen, K.~Bandura, and J.~B. Peterson, ``{Hydrogen 21-cm
  Intensity Mapping at redshift 0.8}'',
  \href{http://dx.doi.org/10.1038/nature09187}{{\em Nature} {\bfseries 466}
  (2010) 463--465}, \href{http://arxiv.org/abs/1007.3709}{{\ttfamily
  arXiv:1007.3709}}.

\bibitem{Kuhlen:2005cm}
M.~Kuhlen, P.~Madau, and R.~Montgomery, ``{The spin temperature and 21cm
  brightness of the intergalactic medium in the pre-reionization era}'',
  \href{http://dx.doi.org/10.1086/500548}{{\em Astrophys. J. Lett.} {\bfseries
  637} (2006) L1--L4}, \href{http://arxiv.org/abs/astro-ph/0510814}{{\ttfamily
  arXiv:astro-ph/0510814}}.

\bibitem{Pillepich:2006fj}
A.~Pillepich, C.~Porciani, and S.~Matarrese, ``{The bispectrum of redshifted
  21-cm fluctuations from the dark ages}'',
  \href{http://dx.doi.org/10.1086/517963}{{\em Astrophys. J.} {\bfseries 662}
  (2007) 1--14}, \href{http://arxiv.org/abs/astro-ph/0611126}{{\ttfamily
  arXiv:astro-ph/0611126}}.

\bibitem{Lewis:2007kz}
A.~Lewis and A.~Challinor, ``{The 21cm angular-power spectrum from the dark
  ages}'', \href{http://dx.doi.org/10.1103/PhysRevD.76.083005}{{\em Phys. Rev.
  D} {\bfseries 76} (2007) 083005},
  \href{http://arxiv.org/abs/astro-ph/0702600}{{\ttfamily
  arXiv:astro-ph/0702600}}.

\bibitem{AliHaimoud2013}
Y.~Ali-Ha\"imoud, P.~D. Meerburg, and S.~Yuan, ``{New light on 21 cm intensity
  fluctuations from the dark ages}'',
  \href{http://dx.doi.org/10.1103/PhysRevD.89.083506}{{\em Phys. Rev. D}
  {\bfseries 89} no.~8, (2014) 083506},
  \href{http://arxiv.org/abs/1312.4948}{{\ttfamily arXiv:1312.4948}}.

\bibitem{Munoz2015}
J.~B. Mu\~noz, Y.~Ali-Ha\"imoud, and M.~Kamionkowski, ``{Primordial
  non-gaussianity from the bispectrum of 21-cm fluctuations in the dark
  ages}'', \href{http://dx.doi.org/10.1103/PhysRevD.92.083508}{{\em Phys. Rev.
  D} {\bfseries 92} no.~8, (2015) 083508},
  \href{http://arxiv.org/abs/1506.04152}{{\ttfamily arXiv:1506.04152}}.

\bibitem{fisher:fishermatrix}
R.~A. Fisher, ``The Fiducial Argument in Statistical Inference'',
  \href{http://dx.doi.org/10.1111/j.1469-1809.1935.tb02120.x}{{\em Annals
  Eugen.} {\bfseries 6} (1935) 391--398}.

\bibitem{bunn:fishermatrix}
E.~F. Bunn, {\em Statistical analysis of cosmic microwave background
  anisotropy}.
\newblock PhD thesis, University of California, Berkeley, 1995.

\bibitem{vogeley:fishermatrix}
M.~S. Vogeley and A.~S. Szalay, ``Eigenmode analysis of galaxy redshift surveys
  I. theory and methods'', \href{http://dx.doi.org/10.1086/177399}{{\em The
  Astrophysical Journal} {\bfseries 465} (1996) 34},
  \href{http://arxiv.org/abs/astro-ph/9601185}{{\ttfamily
  arXiv:astro-ph/9601185}}.

\bibitem{tegmark:fishermatrix}
M.~Tegmark, A.~N. Taylor, and A.~F. Heavens, ``Karhunen-Lo\`eve Eigenvalue
  Problems in Cosmology: How Should We Tackle Large Data Sets?'',
  \href{http://dx.doi.org/10.1086/303939}{{\em The Astrophysical Journal}
  {\bfseries 480} no.~1, (1997) 22},
  \href{http://arxiv.org/abs/astro-ph/9603021}{{\ttfamily
  arXiv:astro-ph/9603021}}.

\bibitem{Bellomo2020}
N.~Bellomo, J.~L. Bernal, G.~Scelfo, A.~Raccanelli, and L.~Verde, ``{Beware of
  commonly used approximations. Part I. Errors in forecasts}'',
  \href{http://dx.doi.org/10.1088/1475-7516/2020/10/016}{{\em JCAP} {\bfseries
  10} (2020) 016}, \href{http://arxiv.org/abs/2005.10384}{{\ttfamily
  arXiv:2005.10384}}.

\bibitem{mozden:edgesskytemperature}
T.~J. {Mozdzen}, J.~D. {Bowman}, R.~A. {Monsalve}, and A.~E.~E. {Rogers},
  ``{Improved measurement of the spectral index of the diffuse radio background
  between 90 and 190 MHz}'',
  \href{http://dx.doi.org/10.1093/mnras/stw2696}{{\em MNRAS} {\bfseries 464}
  no.~4, (Feb, 2017) 4995--5002},
  \href{http://arxiv.org/abs/1609.08705}{{\ttfamily arXiv:1609.08705}}.

\bibitem{rogers:skytemperature}
A.~E.~E. Rogers, J.~D. Bowman, J.~Vierinen, R.~Monsalve, and T.~Mozdzen,
  ``Radiometric measurements of electron temperature and opacity of ionospheric
  perturbations'', \href{http://dx.doi.org/10.1002/2014RS005599}{{\em Radio
  Science} {\bfseries 50} no.~2, (2015) 130--137},
  \href{http://arxiv.org/abs/1412.2255}{{\ttfamily arXiv:1412.2255}}.

\bibitem{liu:dataanalysis21cm}
A.~Liu and J.~R. Shaw, ``Data Analysis for Precision 21 cm Cosmology'',
  \href{http://dx.doi.org/10.1088/1538-3873/ab5bfd}{{\em Publications of the
  Astronomical Society of the Pacific} {\bfseries 132} no.~1012, (Apr, 2020)
  062001}, \href{http://arxiv.org/abs/1907.08211}{{\ttfamily
  arXiv:1907.08211}}.

\bibitem{deoliveiracosta:21cmforegrounds}
A.~De~Oliveira-Costa, M.~Tegmark, B.~M. Gaensler, J.~Jonas, T.~L. Landecker,
  and P.~Reich, ``A model of diffuse Galactic radio emission from 10 MHz to 100
  GHz'', \href{http://dx.doi.org/10.1111/j.1365-2966.2008.13376.x}{{\em Monthly
  Notices of the Royal Astronomical Society} {\bfseries 388} no.~1, (07, 2008)
  247--260}, \href{http://arxiv.org/abs/0802.1525}{{\ttfamily
  arXiv:0802.1525}}.

\bibitem{liu:21cmforegrounds}
A.~Liu and M.~Tegmark, ``How well can we measure and understand foregrounds
  with 21-cm experiments?'',
  \href{http://dx.doi.org/10.1111/j.1365-2966.2011.19989.x}{{\em Monthly
  Notices of the Royal Astronomical Society} {\bfseries 419} no.~4, (01, 2012)
  3491--3504}, \href{http://arxiv.org/abs/1106.0007}{{\ttfamily
  arXiv:1106.0007}}.

\bibitem{zheng:21cmforegrounds}
H.~Zheng {\em et~al.}, ``An improved model of diffuse galactic radio emission
  from 10 MHz to 5 THz'', \href{http://dx.doi.org/10.1093/mnras/stw2525}{{\em
  Monthly Notices of the Royal Astronomical Society} {\bfseries 464} no.~3,
  (10, 2016) 3486--3497}, \href{http://arxiv.org/abs/1605.04920}{{\ttfamily
  arXiv:1605.04920}}.

\bibitem{opperman:foregroundremoval}
{Oppermann, N.} {\em et~al.}, ``An improved map of the Galactic Faraday sky'',
  \href{http://dx.doi.org/10.1051/0004-6361/201118526}{{\em A\&A} {\bfseries
  542} (Jun, 2012) A93}, \href{http://arxiv.org/abs/1111.6186}{{\ttfamily
  arXiv:1111.6186}}.

\bibitem{switzer:foregroundremoval}
E.~R. Switzer and A.~Liu, ``Erasing the variable: empirical foreground
  discovery for global 21-cm spectrum experiments'',
  \href{http://dx.doi.org/10.1088/0004-637X/793/2/102}{{\em The Astrophysical
  Journal} {\bfseries 793} no.~2, (Sep, 2014) 102},
  \href{http://arxiv.org/abs/1404.7561}{{\ttfamily arXiv:1404.7561}}.

\bibitem{wolleben:foregroundremoval}
M.~Wolleben {\em et~al.}, ``The Global Magneto-Ionic Medium Survey: Polarimetry
  of the Southern Sky from 300 to 480 MHz'',
  \href{http://dx.doi.org/10.3847/1538-3881/ab22b0}{{\em The Astronomical
  Journal} {\bfseries 158} no.~1, (Jul, 2019) 44},
  \href{http://arxiv.org/abs/1905.12685}{{\ttfamily arXiv:1905.12685}}.

\bibitem{vedantham:groundobservationlimitationI}
H.~K. Vedantham, L.~V.~E. Koopmans, A.~G. de~Bruyn, S.~J. Wijnholds, B.~Ciardi,
  and M.~A. Brentjens, ``Chromatic effects in the 21-cm global signal from the
  cosmic dawn'', \href{http://dx.doi.org/10.1093/mnras/stt1878}{{\em Monthly
  Notices of the Royal Astronomical Society} {\bfseries 437} no.~2, (11, 2013)
  1056--1069}, \href{http://arxiv.org/abs/1306.2172}{{\ttfamily
  arXiv:1306.2172}}.

\bibitem{datta:groundobservationlimitations}
A.~Datta, R.~Bradley, J.~O. Burns, G.~Harker, A.~Komjathy, and T.~J.~W. Lazio,
  ``The effects of the ionosphere on ground-based detectionof the global 21-cm
  signal from Cosmic Dawn and the Dark Ages'',
  \href{http://dx.doi.org/10.3847/0004-637X/831/1/6}{{\em The Astrophysical
  Journal} {\bfseries 831} no.~1, (Oct, 2016) 6},
  \href{http://arxiv.org/abs/1409.0513}{{\ttfamily arXiv:1409.0513}}.

\bibitem{vedantham:groundobservationlimitationsII}
H.~K. Vedantham and L.~V.~E. Koopmans, ``Scintillation noise in widefield radio
  interferometry'', \href{http://dx.doi.org/10.1093/mnras/stv1594}{{\em Monthly
  Notices of the Royal Astronomical Society} {\bfseries 453} no.~1, (08, 2015)
  925--938}, \href{http://arxiv.org/abs/1412.1420}{{\ttfamily
  arXiv:1412.1420}}.

\bibitem{vedantham:groundobservationlimitationsIII}
H.~K. Vedantham and L.~V.~E. Koopmans, ``Scintillation noise power spectrum and
  its impact on high-redshift 21-cm observations'',
  \href{http://dx.doi.org/10.1093/mnras/stw443}{{\em Monthly Notices of the
  Royal Astronomical Society} {\bfseries 458} no.~3, (02, 2016) 3099--3117},
  \href{http://arxiv.org/abs/1512.00159}{{\ttfamily arXiv:1512.00159}}.

\bibitem{lazio:lunarradioarray}
J.~Lazio, C.~Carilli, J.~Hewitt, S.~Furlanetto, and J.~Burns, ``The lunar radio
  array (LRA)'', in {\em UV/Optical/IR Space Telescopes: Innovative
  Technologies and Concepts IV}, H.~A. MacEwen and J.~B. Breckinridge, eds.,
  vol.~7436, p.~74360I, International Society for Optics and Photonics.
\newblock 2009.
\newblock \url{https://doi.org/10.1117/12.827955}.

\bibitem{datta:foregroundwedge}
A.~Datta, J.~D. Bowman, and C.~L. Carilli, ``Bright source subtraction
  requirements for redshifted 21-cm measurements'',
  \href{http://dx.doi.org/10.1088/0004-637X/724/1/526}{{\em The Astrophysical
  Journal} {\bfseries 724} no.~1, (Nov, 2010) 526},
  \href{http://arxiv.org/abs/1005.4071}{{\ttfamily arXiv:1005.4071}}.

\bibitem{vedantham:foregroundwedge}
H.~Vedantham, N.~U. Shankar, and R.~Subrahmanyan, ``Imaging the epoch of
  reionization: limitations from foreground confusion and imaging algorithms'',
  \href{http://dx.doi.org/10.1088/0004-637X/745/2/176}{{\em The Astrophysical
  Journal} {\bfseries 745} no.~2, (Jan, 2012) 176},
  \href{http://arxiv.org/abs/1106.1297}{{\ttfamily arXiv:1106.1297}}.

\bibitem{morales:foregroundwedge}
M.~F. Morales, B.~Hazelton, I.~Sullivan, and A.~Beardsley, ``Four fundamental
  foreground power spectrum shapes for 21 cm cosmology observations'',
  \href{http://dx.doi.org/10.1088/0004-637X/752/2/137}{{\em The Astrophysical
  Journal} {\bfseries 752} no.~2, (Jun, 2012) 137},
  \href{http://arxiv.org/abs/1202.3830}{{\ttfamily arXiv:1202.3830}}.

\bibitem{parsons:foregroundwedge}
A.~R. Parsons {\em et~al.}, ``A per-baseline, delay-spectrum technique for
  accessing the 21-cm cosmic ionization signature'',
  \href{http://dx.doi.org/10.1088/0004-637X/756/2/165}{{\em The Astrophysical
  Journal} {\bfseries 756} no.~2, (Aug, 2012) 165},
  \href{http://arxiv.org/abs/1204.4749}{{\ttfamily arXiv:1204.4749}}.

\bibitem{trott:foregroundwedge}
C.~M. Trott, R.~B. Wayth, and S.~J. Tingay, ``The impact of point-source
  subtraction residuals on 12-cm epoch of reionization estimation'',
  \href{http://dx.doi.org/10.1088/0004-637X/757/1/101}{{\em The Astrophysical
  Journal} {\bfseries 757} no.~1, (Sep, 2012) 101},
  \href{http://arxiv.org/abs/1208.0646}{{\ttfamily arXiv:1208.0646}}.

\bibitem{hazelton:foregroundwedge}
B.~J. Hazelton, M.~F. Morales, and I.~S. Sullivan, ``The fundamental
  multi-baseline mode-mixing foreground in 21 cm epoch of reionization
  observations'', \href{http://dx.doi.org/10.1088/0004-637X/770/2/156}{{\em The
  Astrophysical Journal} {\bfseries 770} no.~2, (Jun, 2013) 156},
  \href{http://arxiv.org/abs/1301.3126}{{\ttfamily arXiv:1301.3126}}.

\bibitem{pober:foregroundwedge}
J.~C. Pober {\em et~al.}, ``Opening the 21 cm epoch of reionization window:
  measurements of foreground isolation with PAPER'',
  \href{http://dx.doi.org/10.1088/2041-8205/768/2/L36}{{\em The Astrophysical
  Journal Letters} {\bfseries 768} no.~2, (Apr, 2013) L36},
  \href{http://arxiv.org/abs/1301.7099}{{\ttfamily arXiv:1301.7099}}.

\bibitem{liu:foregroundwedgeI}
A.~Liu, A.~R. Parsons, and C.~M. Trott, ``Epoch of reionization window. I.
  Mathematical formalism'',
  \href{http://dx.doi.org/10.1103/PhysRevD.90.023018}{{\em Phys. Rev. D}
  {\bfseries 90} (Jul, 2014) 023018},
  \href{http://arxiv.org/abs/1404.2596}{{\ttfamily arXiv:1404.2596}}.

\bibitem{liu:foregroundwedgeII}
A.~Liu, A.~R. Parsons, and C.~M. Trott, ``Epoch of reionization window. II.
  Statistical methods for foreground wedge reduction'',
  \href{http://dx.doi.org/10.1103/PhysRevD.90.023019}{{\em Phys. Rev. D}
  {\bfseries 90} (Jul, 2014) 023019},
  \href{http://arxiv.org/abs/1404.4372}{{\ttfamily arXiv:1404.4372}}.

\bibitem{Seo:foregroundwedge}
H.-J. Seo and C.~M. Hirata, ``{The foreground wedge and 21 cm BAO surveys}'',
  \href{http://dx.doi.org/10.1093/mnras/stv2806}{{\em Mon. Not. Roy. Astron.
  Soc.} {\bfseries 456} no.~3, (2016) 3142--3156},
  \href{http://arxiv.org/abs/1508.06503}{{\ttfamily arXiv:1508.06503}}.

\bibitem{pober:foregroundwedgesize}
J.~C. Pober {\em et~al.}, ``What next-generation 21-cm power spectrum
  measurements can teach us about the epoch of reionization'',
  \href{http://dx.doi.org/10.1088/0004-637X/782/2/66}{{\em The Astrophysical
  Journal} {\bfseries 782} no.~2, (Jan, 2014) 66},
  \href{http://arxiv.org/abs/1310.7031}{{\ttfamily arXiv:1310.7031}}.

\bibitem{braun:skasensitivity}
R.~{Braun}, A.~{Bonaldi}, T.~{Bourke}, E.~{Keane}, and J.~{Wagg}, ``Anticipated
  Performance of the Square Kilometre Array -- Phase 1 (SKA1)'',
  \href{http://arxiv.org/abs/1912.12699}{{\ttfamily arXiv:1912.12699}}.

\bibitem{macario:skasensitivity}
G.~Macario {\em et~al.}, ``Characterization of the SKA1-Low prototype station
  Aperture Array Verification System 2'',
  \href{http://dx.doi.org/10.1117/1.JATIS.8.1.011014}{{\em Journal of
  Astronomical Telescopes, Instruments, and Systems} {\bfseries 8} no.~1,
  (2022) 011014}, \href{http://arxiv.org/abs/2109.11983}{{\ttfamily
  arXiv:2109.11983}}.

\bibitem{bernal:smbhseeds}
J.~L. Bernal, A.~Raccanelli, L.~Verde, and J.~Silk, ``Signatures of primordial
  black holes as seeds of supermassive black holes'',
  \href{http://dx.doi.org/10.1088/1475-7516/2018/05/017}{{\em Journal of
  Cosmology and Astroparticle Physics} {\bfseries 2018} no.~05, (May, 2018)
  017}, \href{http://arxiv.org/abs/1712.01311}{{\ttfamily arXiv:1712.01311}}.

\bibitem{short:dmdarkages}
K.~Short, J.~L. Bernal, A.~Raccanelli, L.~Verde, and J.~Chluba, ``Enlightening
  the dark ages with dark matter'',
  \href{http://dx.doi.org/10.1088/1475-7516/2020/07/020}{{\em Journal of
  Cosmology and Astroparticle Physics} {\bfseries 2020} no.~07, (Jul, 2020)
  020}, \href{http://arxiv.org/abs/1912.07409}{{\ttfamily arXiv:1912.07409}}.

\bibitem{mcquinn:21cmforecast}
M.~McQuinn, O.~Zahn, M.~Zaldarriaga, L.~Hernquist, and S.~R. Furlanetto,
  ``Cosmological Parameter Estimation Using 21 cm Radiation from the Epoch of
  Reionization'', \href{http://dx.doi.org/10.1086/505167}{{\em The
  Astrophysical Journal} {\bfseries 653} no.~2, (Dec, 2006) 815},
  \href{http://arxiv.org/abs/astro-ph/0512263}{{\ttfamily
  arXiv:astro-ph/0512263}}.

\bibitem{bowman:21cmforecast}
J.~D. Bowman, M.~F. Morales, and J.~N. Hewitt, ``Foreground contamination in
  interferometric measurements of the redshifted 21 cm power spectrum'',
  \href{http://dx.doi.org/10.1088/0004-637X/695/1/183}{{\em The Astrophysical
  Journal} {\bfseries 695} no.~1, (Mar, 2009) 183},
  \href{http://arxiv.org/abs/0807.3956}{{\ttfamily arXiv:0807.3956}}.

\bibitem{Blas:2011rf}
D.~Blas, J.~Lesgourgues, and T.~Tram, ``{The Cosmic Linear Anisotropy Solving
  System (CLASS) II: Approximation schemes}'',
  \href{http://dx.doi.org/10.1088/1475-7516/2011/07/034}{{\em JCAP} {\bfseries
  07} (2011) 034}, \href{http://arxiv.org/abs/1104.2933}{{\ttfamily
  arXiv:1104.2933}}.

\bibitem{Amendola:2016saw}
L.~Amendola {\em et~al.}, ``{Cosmology and fundamental physics with the Euclid
  satellite}'', \href{http://dx.doi.org/10.1007/s41114-017-0010-3}{{\em Living
  Rev. Rel.} {\bfseries 21} no.~1, (2018) 2},
  \href{http://arxiv.org/abs/1606.00180}{{\ttfamily arXiv:1606.00180}}.

\bibitem{ade:planckmodifiedgravity}
The {\bfseries Planck Collaboration}, P.~A.~R. Ade {\em et~al.}, ``Planck 2015
  results - XIV. Dark energy and modified gravity'',
  \href{http://dx.doi.org/10.1051/0004-6361/201525814}{{\em A\&A} {\bfseries
  594} (2016) A14}, \href{http://arxiv.org/abs/1502.01590}{{\ttfamily
  arXiv:1502.01590}}.

\bibitem{Durrer:2008in}
R.~Durrer and R.~Maartens, ``{Dark Energy and Modified Gravity}'',
\newblock 11, 2008.
\newblock \href{http://arxiv.org/abs/0811.4132}{{\ttfamily arXiv:0811.4132}}.

\bibitem{amendola:mgparametrization}
L.~Amendola, M.~Kunz, and D.~Sapone, ``Measuring the dark side (with weak
  lensing)'', \href{http://dx.doi.org/10.1088/1475-7516/2008/04/013}{{\em
  Journal of Cosmology and Astroparticle Physics} {\bfseries 2008} no.~04,
  (Apr, 2008) 013}, \href{http://arxiv.org/abs/0704.2421}{{\ttfamily
  arXiv:0704.2421}}.

\bibitem{zhao:muetaparametrization}
G.-B. Zhao, T.~Giannantonio, L.~Pogosian, A.~Silvestri, D.~J. Bacon, K.~Koyama,
  R.~C. Nichol, and Y.-S. Song, ``Probing modifications of general relativity
  using current cosmological observations'',
  \href{http://dx.doi.org/10.1103/PhysRevD.81.103510}{{\em Phys. Rev. D}
  {\bfseries 81} (May, 2010) 103510},
  \href{http://arxiv.org/abs/1003.0001}{{\ttfamily arXiv:1003.0001}}.

\bibitem{alonso:screening}
D.~Alonso, E.~Bellini, P.~G. Ferreira, and M.~Zumalac\'arregui, ``Observational
  future of cosmological scalar-tensor theories'',
  \href{http://dx.doi.org/10.1103/PhysRevD.95.063502}{{\em Phys. Rev. D}
  {\bfseries 95} (Mar, 2017) 063502},
  \href{http://arxiv.org/abs/1610.09290}{{\ttfamily arXiv:1610.09290}}.

\bibitem{spuriomancini:screening}
A.~Spurio Mancini, R.~Reischke, V.~Pettorino, B.~M. Sch\"afer, and
  M.~Zumalac\'arregui, ``Testing (modified) gravity with 3D and tomographic
  cosmic shear'', \href{http://dx.doi.org/10.1093/mnras/sty2092}{{\em Monthly
  Notices of the Royal Astronomical Society} {\bfseries 480} no.~3, (08, 2018)
  3725--3738}, \href{http://arxiv.org/abs/1801.04251}{{\ttfamily
  arXiv:1801.04251}}.

\bibitem{bosi:gwxlss}
M.~Bosi, N.~Bellomo, and A.~Raccanelli, ``Constraining extended cosmologies
  with GW×LSS cross-correlations'',
  \href{http://dx.doi.org/10.1088/1475-7516/2023/11/086}{{\em Journal of
  Cosmology and Astroparticle Physics} {\bfseries 2023} no.~11, (Nov, 2023)
  086}, \href{http://arxiv.org/abs/2306.03031}{{\ttfamily arXiv:2306.03031}}.

\bibitem{bellini:eftofde}
E.~Bellini and I.~Sawicki, ``Maximal freedom at minimum cost: linear
  large-scale structure in general modifications of gravity'',
  \href{http://dx.doi.org/10.1088/1475-7516/2014/07/050}{{\em Journal of
  Cosmology and Astroparticle Physics} {\bfseries 2014} no.~07, (Jul, 2014)
  050}, \href{http://arxiv.org/abs/1404.3713}{{\ttfamily arXiv:1404.3713}}.

\bibitem{khoury:chameleonI}
J.~Khoury and A.~Weltman, ``Chameleon Fields: Awaiting Surprises for Tests of
  Gravity in Space'',
  \href{http://dx.doi.org/10.1103/PhysRevLett.93.171104}{{\em Phys. Rev. Lett.}
  {\bfseries 93} (Oct, 2004) 171104},
  \href{http://arxiv.org/abs/astro-ph/0309300}{{\ttfamily
  arXiv:astro-ph/0309300}}.

\bibitem{khoury:chameleonII}
J.~Khoury and A.~Weltman, ``Chameleon cosmology'',
  \href{http://dx.doi.org/10.1103/PhysRevD.69.044026}{{\em Phys. Rev. D}
  {\bfseries 69} (Feb, 2004) 044026},
  \href{http://arxiv.org/abs/astro-ph/0309411}{{\ttfamily
  arXiv:astro-ph/0309411}}.

\bibitem{hinterbichler:symmetron}
K.~Hinterbichler and J.~Khoury, ``Screening Long-Range Forces through Local
  Symmetry Restoration'',
  \href{http://dx.doi.org/10.1103/PhysRevLett.104.231301}{{\em Phys. Rev.
  Lett.} {\bfseries 104} (Jun, 2010) 231301},
  \href{http://arxiv.org/abs/1001.4525}{{\ttfamily arXiv:1001.4525}}.

\bibitem{brax:dilaton}
P.~Brax, C.~van~de Bruck, A.-C. Davis, B.~Li, and D.~J. Shaw, ``Nonlinear
  structure formation with the environmentally dependent dilaton'',
  \href{http://dx.doi.org/10.1103/PhysRevD.83.104026}{{\em Phys. Rev. D}
  {\bfseries 83} (May, 2011) 104026},
  \href{http://arxiv.org/abs/1102.3692}{{\ttfamily arXiv:1102.3692}}.

\bibitem{babichev:kmouflage}
E.~Babichev, C.~Deffayet, and R.~Ziour, ``k-Mouflage gravity'',
  \href{http://dx.doi.org/10.1142/S0218271809016107}{{\em International Journal
  of Modern Physics D} {\bfseries 18} no.~14, (2009) 2147--2154},
  \href{http://arxiv.org/abs/0905.2943}{{\ttfamily arXiv:0905.2943}}.

\bibitem{vainshtein:vainshtein}
A.~Vainshtein, ``To the problem of nonvanishing gravitation mass'',
  \href{http://dx.doi.org/10.1016/0370-2693(72)90147-5}{{\em Physics Letters B}
  {\bfseries 39} no.~3, (1972) 393--394}.

\bibitem{Gregory:2007xy}
R.~Gregory, N.~Kaloper, R.~C. Myers, and A.~Padilla, ``{A New perspective on
  DGP gravity}'', \href{http://dx.doi.org/10.1088/1126-6708/2007/10/069}{{\em
  JHEP} {\bfseries 10} (2007) 069},
  \href{http://arxiv.org/abs/0707.2666}{{\ttfamily arXiv:0707.2666}}.

\bibitem{koyama:mudgpmodel}
K.~Koyama and R.~Maartens, ``Structure formation in the
  Dvali–Gabadadze–Porrati cosmological model'',
  \href{http://dx.doi.org/10.1088/1475-7516/2006/01/016}{{\em Journal of
  Cosmology and Astroparticle Physics} {\bfseries 2006} no.~01, (Jan, 2006)
  016}, \href{http://arxiv.org/abs/astro-ph/0511634}{{\ttfamily
  arXiv:astro-ph/0511634}}.

\bibitem{Schmidt:2009sv}
F.~Schmidt, ``{Cosmological Simulations of Normal-Branch Braneworld Gravity}'',
  \href{http://dx.doi.org/10.1103/PhysRevD.80.123003}{{\em Phys. Rev. D}
  {\bfseries 80} (2009) 123003},
  \href{http://arxiv.org/abs/0910.0235}{{\ttfamily arXiv:0910.0235}}.

\bibitem{lue:mudgpmodel}
A.~Lue, R.~Scoccimarro, and G.~D. Starkman, ``Probing Newton's constant on vast
  scales: Dvali-Gabadadze-Porrati gravity, cosmic acceleration, and large scale
  structure'', \href{http://dx.doi.org/10.1103/PhysRevD.69.124015}{{\em Phys.
  Rev. D} {\bfseries 69} (Jun, 2004) 124015},
  \href{http://arxiv.org/abs/astro-ph/0401515}{{\ttfamily
  arXiv:astro-ph/0401515}}.

\bibitem{Piga:2022mge}
L.~Piga, M.~Marinucci, G.~D'Amico, M.~Pietroni, F.~Vernizzi, and B.~S. Wright,
  ``Constraints on modified gravity from the BOSS galaxy survey'',
  \href{http://dx.doi.org/10.1088/1475-7516/2023/04/038}{{\em JCAP} {\bfseries
  04} (2023) 038}, \href{http://arxiv.org/abs/2211.12523}{{\ttfamily
  arXiv:2211.12523}}.

\bibitem{Karwal:2016vyq}
T.~Karwal and M.~Kamionkowski, ``{Dark energy at early times, the Hubble
  parameter, and the string axiverse}'',
  \href{http://dx.doi.org/10.1103/PhysRevD.94.103523}{{\em Phys. Rev. D}
  {\bfseries 94} no.~10, (2016) 103523},
  \href{http://arxiv.org/abs/1608.01309}{{\ttfamily arXiv:1608.01309}}.

\bibitem{Poulin:2018cxd}
V.~Poulin, T.~L. Smith, T.~Karwal, and M.~Kamionkowski, ``{Early Dark Energy
  Can Resolve The Hubble Tension}'',
  \href{http://dx.doi.org/10.1103/PhysRevLett.122.221301}{{\em Phys. Rev.
  Lett.} {\bfseries 122} no.~22, (2019) 221301},
  \href{http://arxiv.org/abs/1811.04083}{{\ttfamily arXiv:1811.04083}}.

\bibitem{Lin:2019qug}
M.-X. Lin, G.~Benevento, W.~Hu, and M.~Raveri, ``{Acoustic Dark Energy:
  Potential Conversion of the Hubble Tension}'',
  \href{http://dx.doi.org/10.1103/PhysRevD.100.063542}{{\em Phys. Rev. D}
  {\bfseries 100} no.~6, (2019) 063542},
  \href{http://arxiv.org/abs/1905.12618}{{\ttfamily arXiv:1905.12618}}.

\bibitem{Smith:2019ihp}
T.~L. Smith, V.~Poulin, and M.~A. Amin, ``{Oscillating scalar fields and the
  Hubble tension: a resolution with novel signatures}'',
  \href{http://dx.doi.org/10.1103/PhysRevD.101.063523}{{\em Phys. Rev. D}
  {\bfseries 101} no.~6, (2020) 063523},
  \href{http://arxiv.org/abs/1908.06995}{{\ttfamily arXiv:1908.06995}}.

\bibitem{sobotka:2024tat}
A.~C. Sobotka, A.~L. Erickcek, and T.~L. Smith, ``{Signatures of very early
  dark energy in the matter power spectrum}'',
  \href{http://dx.doi.org/10.1103/9bd9-fzwh}{{\em Phys. Rev. D} {\bfseries 111}
  no.~12, (2025) 123522}, \href{http://arxiv.org/abs/2409.06778}{{\ttfamily
  arXiv:2409.06778}}.

\bibitem{Marsh:2011gr}
D.~J.~E. Marsh, ``{The Axiverse Extended: Vacuum Destabilisation, Early Dark
  Energy and Cosmological Collapse}'',
  \href{http://dx.doi.org/10.1103/PhysRevD.83.123526}{{\em Phys. Rev. D}
  {\bfseries 83} (2011) 123526},
  \href{http://arxiv.org/abs/1102.4851}{{\ttfamily arXiv:1102.4851}}.

\bibitem{Hlozek:2014lca}
R.~Hlozek, D.~Grin, D.~J.~E. Marsh, and P.~G. Ferreira, ``{A search for
  ultralight axions using precision cosmological data}'',
  \href{http://dx.doi.org/10.1103/PhysRevD.91.103512}{{\em Phys. Rev. D}
  {\bfseries 91} no.~10, (2015) 103512},
  \href{http://arxiv.org/abs/1410.2896}{{\ttfamily arXiv:1410.2896}}.

\bibitem{Marsh:2015xka}
D.~J.~E. Marsh, ``{Axion Cosmology}'',
  \href{http://dx.doi.org/10.1016/j.physrep.2016.06.005}{{\em Phys. Rept.}
  {\bfseries 643} (2016) 1--79},
  \href{http://arxiv.org/abs/1510.07633}{{\ttfamily arXiv:1510.07633}}.

\bibitem{Hill:2020osr}
J.~C. Hill, E.~McDonough, M.~W. Toomey, and S.~Alexander, ``{Early dark energy
  does not restore cosmological concordance}'',
  \href{http://dx.doi.org/10.1103/PhysRevD.102.043507}{{\em Phys. Rev. D}
  {\bfseries 102} no.~4, (2020) 043507},
  \href{http://arxiv.org/abs/2003.07355}{{\ttfamily arXiv:2003.07355}}.

\bibitem{avila:growthrateoverview}
F.~Avila, A.~Bernui, A.~Bonilla, and R.~C. Nunes, ``Inferring $S_8(z)$ and
  $\gamma (z)$ with cosmic growth rate measurements using machine learning'',
  \href{http://dx.doi.org/10.1140/epjc/s10052-022-10561-0}{{\em Eur. Phys. J.
  C} {\bfseries 82} no.~7, (2022) 594},
  \href{http://arxiv.org/abs/2201.07829}{{\ttfamily arXiv:2201.07829}}.

\bibitem{dore:spherexwhitepaperI}
O.~Dor\'e {\em et~al.}, ``Cosmology with the SPHEREX All-Sky Spectral Survey'',
  \href{http://arxiv.org/abs/1412.4872}{{\ttfamily arXiv:1412.4872}}.

\bibitem{dore:spherexwhitepaperII}
O.~Dor\'e {\em et~al.}, ``Science Impacts of the SPHEREx All-Sky Optical to
  Near-Infrared Spectral Survey: Report of a Community Workshop Examining
  Extragalactic, Galactic, Stellar and Planetary Science'',
  \href{http://arxiv.org/abs/1606.07039}{{\ttfamily arXiv:1606.07039}}.

\bibitem{dore:spherexwhitepaperIII}
O.~Dor\'e {\em et~al.}, ``Science Impacts of the SPHEREx All-Sky Optical to
  Near-Infrared Spectral Survey II: Report of a Community Workshop on the
  Scientific Synergies Between the SPHEREx Survey and Other Astronomy
  Observatories'', \href{http://arxiv.org/abs/1805.05489}{{\ttfamily
  arXiv:1805.05489}}.

\bibitem{Spergel2015:Roman}
D.~Spergel {\em et~al.}, ``{Wide-Field InfrarRed Survey Telescope-Astrophysics
  Focused Telescope Assets WFIRST-AFTA 2015 Report}'',
  \href{http://arxiv.org/abs/1503.03757}{{\ttfamily arXiv:1503.03757}}.

\bibitem{Schlegel:2019eqc}
D.~J. Schlegel {\em et~al.}, ``{Astro2020 APC White Paper: The MegaMapper: a z
  \ensuremath{>} 2 Spectroscopic Instrument for the Study of Inflation and Dark
  Energy}'', {\em Bull. Am. Astron. Soc.} {\bfseries 51} no.~7, (2019) 229,
  \href{http://arxiv.org/abs/1907.11171}{{\ttfamily arXiv:1907.11171}}.

\bibitem{Schlegel:2022vrv}
D.~J. Schlegel {\em et~al.}, ``{The MegaMapper: A Stage-5 Spectroscopic
  Instrument Concept for the Study of Inflation and Dark Energy}'',
  \href{http://arxiv.org/abs/2209.04322}{{\ttfamily arXiv:2209.04322}}.

\bibitem{SIRMOS}
R.~Content {\em et~al.}, \href{http://dx.doi.org/10.1117/12.3017865}{``{SIRMOS:
  NIR spectroscopy of 131,000,000 galaxies over $1 < z < 4$ and R~1300}'',} in
  {\em Space Telescopes and Instrumentation 2024: Optical, Infrared, and
  Millimeter Wave}, L.~E. Coyle, S.~Matsuura, and M.~D. Perrin, eds.,
  vol.~13092, p.~130920Z, International Society for Optics and Photonics.
\newblock 2024.

\bibitem{frusciante:nDGPeuclidforecast}
The {\bfseries Euclid Collaboration}, N.~Frusciante {\em et~al.}, ``Euclid:
  Constraining linearly scale-independent modifications of gravity with the
  spectroscopic and photometric primary probes'',
  \href{http://dx.doi.org/10.1051/0004-6361/202347526}{{\em A\&A} {\bfseries
  690} (2024) A133}, \href{http://arxiv.org/abs/2306.12368}{{\ttfamily
  arXiv:2306.12368}}.

\bibitem{bose:nDGPeuclidforecast}
The {\bfseries Euclid Collaboration}, B.~Bose {\em et~al.}, ``Euclid
  preparation - XLIV. Modelling spectroscopic clustering on mildly nonlinear
  scales in beyond-$\Lambda$CDM models'',
  \href{http://dx.doi.org/10.1051/0004-6361/202348784}{{\em A\&A} {\bfseries
  689} (2024) A275}, \href{http://arxiv.org/abs/2311.13529}{{\ttfamily
  arXiv:2311.13529}}.

\bibitem{McDonough:2023qcu}
E.~McDonough, J.~C. Hill, M.~M. Ivanov, A.~La~Posta, and M.~W. Toomey,
  ``{Observational constraints on early dark energy}'',
  \href{http://dx.doi.org/10.1142/S0218271824300039}{{\em Int. J. Mod. Phys. D}
  {\bfseries 33} no.~11, (2024) 2430003},
  \href{http://arxiv.org/abs/2310.19899}{{\ttfamily arXiv:2310.19899}}.

\bibitem{venumadhav:21cmlinefinitewidth}
T.~Venumadhav, L.~Dai, A.~Kaurov, and M.~Zaldarriaga, ``Heating of the
  intergalactic medium by the cosmic microwave background during cosmic dawn'',
  \href{http://dx.doi.org/10.1103/PhysRevD.98.103513}{{\em Phys. Rev. D}
  {\bfseries 98} (Nov, 2018) 103513},
  \href{http://arxiv.org/abs/1804.02406}{{\ttfamily arXiv:1804.02406}}.

\end{thebibliography}\endgroup
\bibliographystyle{utcaps}

\end{document}